\documentclass[,aps,physrev,preprint,superscriptaddress,longbibliography]{revtex4-2}

\usepackage{graphicx} 
\usepackage{float}
\usepackage{amsmath,amssymb}

\usepackage{url}
\usepackage[colorlinks=true,  urlcolor=blue, linkcolor=black, citecolor=blue]{hyperref}

\usepackage{subfig}
\usepackage{rotating}    

\usepackage{comment}

\usepackage{tikz}
\usepackage{xcolor,soul}

\newcommand{\micheline}[1]{[\textcolor{orange}{M: \textbf{#1}}]}

\begin{document}

\title{Fluid flow in low aspect-ratio curved channels: from small to moderate Dean numbers. }


\author{Ezzahrae Jaafari}%
\affiliation{Université de Toulouse, Toulouse INP, CNRS, Laboratoire de Génie Chimique (LGC), Toulouse, France}

\author{Pascale Magaud}%
\affiliation{Institut Clément Ader (ICA), Université de Toulouse CNRS, INSA, ISAE-SUPAERO, IMT Mines-Albi, Toulouse, France}
\affiliation{Université de Toulouse, CNRS, Toulouse INP, INSA, Fédération FERMaT, Toulouse, France}

\author{Micheline Abbas}%
\affiliation{Université de Toulouse, Toulouse INP, CNRS, Laboratoire de Génie Chimique (LGC), Toulouse, France}
\affiliation{Université de Toulouse, CNRS, Toulouse INP, INSA, Fédération FERMaT, Toulouse, France}
 \email{micheline.abbas@toulouse-inp.fr}




\begin{abstract}
The pressure-driven flow is numerically investigated in curved channels at low aspect ratio, where centrifugal forces act along the largest dimension. 
The dynamics is studied numerically, as a function of the characteristic Dean number, $\mathrm{De}=\mathrm{Re}\sqrt{\delta}$, by varying independently the Reynolds number $\mathrm{Re}$ and the curvature ratio $\delta$, the ratio between the hydraulic diameter and the radius of curvature. A wide range of dimensionless numbers is considered; $\mathrm{De}\lesssim200$ and $0.005\leq\delta\leq0.15$.
For $\mathrm{De}\lesssim 100$, the flow remains steady, whereas at larger Dean numbers, the flow is stable for several turns before transient structures developed. While investigating the flow features in the stable regime, only one pair of counter-rotating vortices is observed. 
At small $\mathrm{De}$ and large $\delta$, the peak of the streamwise velocity and the center of the vortices are located near the inner channel wall. They both shift toward the outer wall as $\mathrm{De}$ is increased and/or $\delta$ is decreased, a feature that is expected to affect the transport of particles in curved channels. A scaling law for the secondary flow is formulated from dimensional analysis, rather than relying on empirical correlations. The friction coefficient of the flow as well as the development angle are also rationalized in terms of both $\mathrm{Re}$ and $\delta$. 
\end{abstract}


\maketitle

\section{Introduction}

Flow in curved channels is of particular interest for particle sorting and concentration, as the interplay between inertia and curvature promotes particle focusing \cite{martel2013particle}. Low-aspect-ratio channels are investigated in this study, motivated by spiral microfluidic experiments that demonstrated efficient focusing and separation of both inert and biological particles at finite particle Reynolds numbers \cite{howell2025experimental, hill2022efficient}. 
A detailed understanding of particle dynamics in such systems requires careful characterization of the single-phase flow, which remains incompletely addressed.

Flow in curved geometries exhibits secondary flows even in the laminar regime. Early insights date back to Thomson in the 19th century \cite{thomson1877}, while Dean (1928) \cite{dean1928} established the first theoretical framework for fully developed pressure-driven flow in curved channels using a perturbation approach to the Navier–Stokes equations. He showed that secondary flows arise as a pair of counter-rotating vortices in the cross-section, for both circular and rectangular geometries. The resulting velocity field depends on a dimensionless parameter—later termed the Dean number—defined as the product of the Reynolds number and the square root of the curvature ratio. In the laminar regime, the flow structure further depends on the cross-sectional shape \cite{Cuming1952} and, for non-Newtonian fluids, on rheological properties \cite{thomas1965}. Subsequent numerical studies \cite{austin1973, joseph_smith1975, cheng1976} examined fully developed flow in curved rectangular channels, while combined experimental and computational investigations addressed developing flow in curved ducts over a range of Dean numbers \cite{goldstein1967, ghia1977, humphrey1977, hille1985}. A comprehensive review of laminar incompressible flow in curved pipes was later provided by Berger \emph{et al.} (1983) \cite{berger1983}. 

In curved rectangular channels, when the Dean number exceeds a critical value $\mathrm{De}_c$, an additional pair of vortices emerges in the cross-section due to an imbalance between radial pressure gradient and centrifugal forces near the outer wall \cite{cheng1976}. This transition from a two-cell to a four-cell structure has been extensively studied, notably by Winters (1987) \cite{winters1987}, who conducted a detailed bifurcation analysis in square ducts. Subsequent works \cite{daskopoulos1989, kao1992, mees1996} extended the analysis to various cross-sectional geometries and a wide range of Dean numbers, reporting even transient six-cell flow states at high Dean numbers. These studies demonstrate that the critical Dean number 
$\mathrm{De}_c$, which triggers the transition from two- to four- or even six-cells, depends on the channel aspect ratio  (see, e.g., Fig. 10 in \cite{Fellouah2006}). 
\textcolor{black}{The works that considered flow in channels with low aspect ratio $\lambda$, are sparse. The aspect ratio is defined here as the ratio between channel height $b$ (along the axis) and width $a$ (along the radial direction). 
We consider the definition of the Dean number as $De=\frac{U_0d_h}{\nu}\sqrt{\frac{d_h}{2R}}$, where $\nu$ and $U_0$ denote the fluid kinematic viscosity and average velocity, $d_h$ the channel hydraulic diameter and $R$ the mean radius of curvature. 
Recent experiments by Nivedita \emph{et al.} \cite{Nivedita2017} reported the onset of four-cells (two vortex pairs) at relatively low Dean numbers, 
$De_c \sim 40$ at low aspect ratio, $\lambda=0.2$. This contrasts with the empirical law $\lambda^{1/2} \mathrm{De}_c \sim 110$ found by Kim \emph{et al.} \cite{Kim2023}, from numerical simulations, and with the law found by Gauthier et al. \cite{gauthier2001centrifugal} $De_c\approx 100$, , from experiments (their definition of the Dean number is slightly different, $De=\frac{ U_0b}{\nu}\sqrt{\frac{a}{2R}}$). The origin of this discrepancy remains unclear.}

\textcolor{black}{Beyond their structure, the intensity of secondary flows is crucial for predicting transport in curved channels. While several studies have reported velocity profiles of the primary and secondary flows \cite{mees1996,Fellouah2006,Kim2023}, only a few have examined how their intensity depends on system parameters. Bayat and Rezai \cite{bayat2017semi}, for example, proposed an empirical power law based on experiments at $\lambda=0.5$ and a single curvature ratio, while Harding \cite{harding2022new} derived a scaling law from the asymptotic low-Reynolds-number regime. Overall, physically grounded scaling laws for secondary-flow intensity remain lacking.}

\textcolor{black}{In this work, we investigate the fluid dynamics in a curved channel with small aspect ratio, fixed to 
$\lambda=\frac{3}{17}$ unless otherwise stated, as motivated by recent experiments in spiral microchannels \cite{hill2022efficient, howell2025experimental, capet6515712microalgae}. 
Our study spans a wide range of curvatures and Reynolds numbers, and examines the primary and secondary flows, the wall friction and the entry length providing a complete self-consistent description of the dynamics. Based on dimensional analysis, two distinct regimes are identified corresponding to cases where centrifugal and viscous forces dominate and another where inertial and centrifugal forces become dominant. Thus, we highlight the correct behaviour of secondary flow intensity, in contrast to empirical power laws that persist through the literature.}

The paper is organized as follows. The flow equations are written in dimensionless form in Section \ref{sec:Dimensional}, allowing to discuss the contribution of inertial, centrifugal and viscous terms to the main and secondary flows as a function of the Reynolds number and the channel curvature. The numerical method is then succinctly presented in Section \ref{sec:num_meth}. The primary and secondary flow are then characterized in Section \ref{sec:primary_flow} and \ref{sec:secondary_flow}. Then, Sections \ref{sec:friction_factor} and \ref{sec:entry_length} discuss the dimensionless friction factor and the entry length necessary for flow development, respectively, showing their scaling laws from small to moderate Dean numbers. The paper ends with concluding remarks. 

\section{Dimensional analysis} \label{sec:Dimensional}

The fluid equations of motion satisfy the mass and and momentum conservation, assuming the fluid flow is incompressible : 

\begin{subequations} 
\begin{align}
\nabla \cdot \mathbf{u} &= 0, \\
 \dfrac{\partial \mathbf{u}}{\partial t} 
+ (\mathbf{u} \cdot \nabla) \mathbf{u} 
&= - \frac{1}{\rho} \nabla p
+\nu \nabla \cdot \Big(  \nabla \mathbf{u} + \nabla \mathbf{u}^T \Big)
\end{align}
\label{eq:NS_system}
\end{subequations}
\noindent where \(\rho\) and \(\nu\) denote the fluid density and dynamic viscosity, while $\textbf{u}$ and $p$ refer to the velocity and pressure fields, respectively. 

\begin{figure} [ht]
    \centering
    \includegraphics[width=0.65\linewidth]{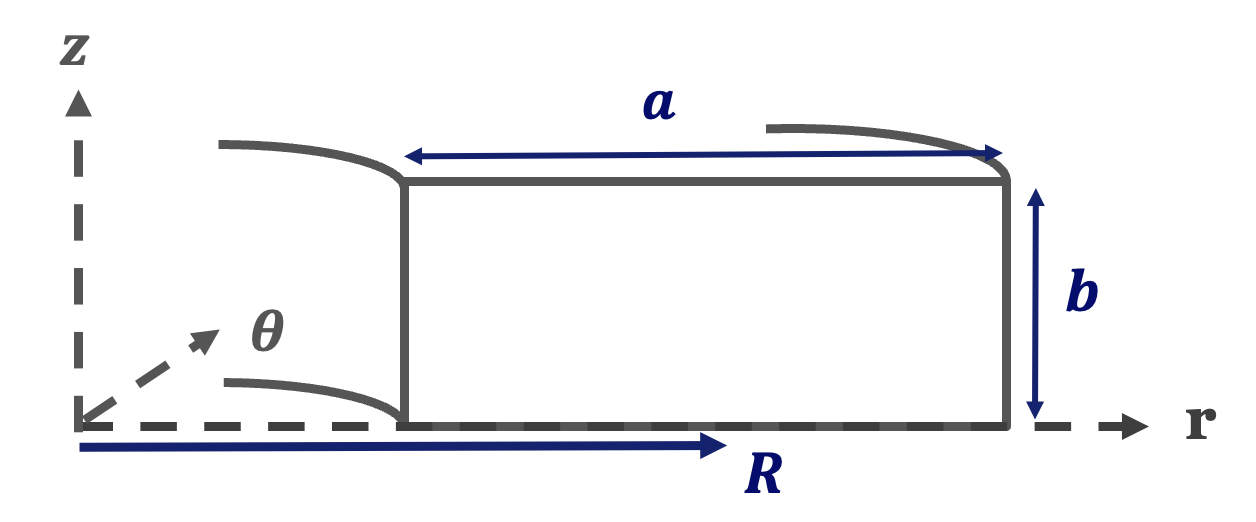}
    \caption{Schematic representation of the flow geometry and the coordinate system.}
    \label{fig:Curved channel geometry}
\end{figure}

To continue further, the equations are written in the Galilean frame of reference, using the cylindrical coordinates, shown in \autoref{fig:Curved channel geometry}. $z$, $r$ and $\theta$ denote the axial, radial and tangential coordinates, respectively. The study is focused on channels with a rectangular cross-section of low aspect ratio, $ \lambda=\dfrac{b}{a} <1 $, where $b$ and $a$ refer to the channel height and width in the axial and radial directions, respectively. The radius of curvature $R$ denotes the distance between the axis and the centerline of the channel, and it is assumed constant in this work. 
To characterize the curvature, we use the curvature ratio $\delta$ : 

\[
\begin{array}{llll}
\delta =\dfrac{d_h}{2R}; 
\end{array}
\]

\noindent with $d_h=\dfrac{2ab}{a+b}$ being the hydraulic diameter. In addition, we introduce the ratio $D=\dfrac{d_h}{a}$, in which the hydraulic diameter is scaled by the channel width. In the limit $\lambda\rightarrow0$, the hydraulic diameter approaches $2b$ and therefore $D\rightarrow2\lambda$. In order to write the flow equations of motion in a dimensionless way, the spatial coordinates, the velocity components $v_r$, $v_\theta$ and $v_x$, as well as the pressure field $p$ are scaled as following:  
\[
\begin{array}{llll}

\overline{r} = \dfrac{r-R}{a}; & \overline{z} = \dfrac{z}{b}; &  &    \\[2ex]

\overline{v}_r = \dfrac{v_r}{U_0}; & \overline{v}_\theta = \dfrac{v_\theta}{U_0}; & \overline{v}_z = \dfrac{v_z}{U_0}; & \overline{p} = \dfrac{p}{\rho U_0^2}
\end{array}
\]
\noindent where $U_0$ denotes the \textcolor{black}{average streamwise velocity}. \\

The flow will be fully characterized using the Reynolds number $\mathrm{Re}$ as usual in pipe flows, and the Dean number $\mathrm{De}$ to account for curvature. The Dean number is interpreted as the ratio between the square root of the inertial and centrifugal forces over the viscous force. Various definitions of the Dean number can be found in the literature, most of them being compiled by Berger \emph{et al.} \cite{berger1983} and more recently by Saffar \emph{et al.} \cite{Yeganeh2023} in the frame of microfluidic applications. In the present study, the following definition is used: 

\[
\begin{array}{llll}
\mathrm{Re}=\dfrac{\rho U_0 d_h}{\mu}; & \mathrm{De}=\mathrm{Re} \sqrt{\delta}; & & \\
\end{array}
\]

\noindent The dimensionless mass and momentum conservation equations become: 

\begin{subequations} 
\begin{align}
\dfrac{\lambda}{r^*} \dfrac{\partial}{\partial \overline{r}}\left( \ r^* \ \overline{v}_r \right) + \dfrac{\partial \overline{v}_z}{\partial \overline{z}} &= 0 \\
\overline{v}_r \dfrac{\partial \overline{v}_r}{\partial \overline{r}}  - \dfrac{2}{D} \dfrac{\delta}{r^*} \overline{v}_\theta^2 + \dfrac{\overline{v}_z}{\lambda} \dfrac{\partial \overline{v}_r}{\partial \overline{z}} 
&= -\dfrac{\partial \overline{p}}{\partial \overline{r}} + \dfrac{1}{Re} \left[ 2 \delta \dfrac{\partial}{\partial \overline{r}} \left( \dfrac{\overline{v}_r}{r^*} \right) +D \dfrac{\partial^2 \overline{v}_r}{\partial \overline{r}^2} +\dfrac{D}{\lambda^2}\dfrac{\partial^2 \overline{v}_r}{\partial \overline{z}^2} \right] \\
\overline{v}_r \dfrac{\partial \overline{v}_\theta}{\partial \overline{r}}  + \dfrac{2}{D} \dfrac{\delta}{r^*} \overline{v}_r \overline{v}_\theta + \dfrac{\overline{v}_z}{\lambda} \dfrac{\partial \overline{v}_\theta}{\partial \overline{z}} 
&= - \dfrac{1}{r^*}\dfrac{\partial \overline{p}}{\partial \theta} + \dfrac{1}{Re} \left[ 2 \delta \dfrac{\partial}{\partial \overline{r}} \left( \dfrac{\overline{v}_\theta}{r^*} \right) +D \dfrac{\partial^2 \overline{v}_\theta}{\partial \overline{r}^2} +\dfrac{D}{\lambda^2}\dfrac{\partial^2 \overline{v}_\theta}{\partial \overline{z}^2} \right]  \\
\overline{v}_r \dfrac{\partial \overline{v}_z}{\partial \overline{r}}  + \dfrac{\overline{v}_z}{\lambda} \dfrac{\partial \overline{v}_z}{\partial \overline{z}} 
&= - \dfrac{1}{\lambda}\dfrac{\partial \overline{p}}{\partial \overline{z}} + \dfrac{1}{Re} \left[ D \dfrac{\partial^2 \overline{v}_z}{\partial \overline{r}^2} 
+ \dfrac{2 \delta}{r^*} \dfrac{\partial \overline{v}_z}{\partial \overline{r}} + \dfrac{D}{\lambda^2}\dfrac{\partial^2 \overline{v}_z}{\partial \overline{z}^2} \right]
\end{align}
\label{eq:NS_cyl}
\end{subequations}

\noindent where $r^* = (1+2 \dfrac{\delta}{D}\overline{r}) \simeq 1$ when the ratio $ \dfrac{\delta}{D}$ is small (for large radius of curvature). \\

\textcolor{black}{At low Reynolds number (and therefore low Dean number and $\delta\ll 1$), the inertial terms in eqs. \ref{eq:NS_cyl} are small. In the radial direction, this results in a balance between the viscous, pressure-gradient and centrifugal terms. Assuming the viscous term balances the centrifugal term, the radial dimensionless velocity $\overline{v}_r$ scales like $\left(\dfrac{\lambda^2}{D^2}Re\delta\right)$. At high Reynolds number (and therefore moderate Dean number), the viscous terms have negligible contribution outside the boundary layer near the walls. Considering that the inertial and centrifugal terms balance each other leads the radial velocity $\overline{v}_r$ to scale like $\left(\sqrt{\dfrac{\delta}{D}}\right)$ .}

\section{Numerical method} \label{sec:num_meth}

The numerical simulations of single-phase flow in a curved channel were carried out with the in-house JADIM code \cite{calmet1997}\cite{Magnaudet1995}, designed to solve the three dimensional unsteady Navier-Stokes equations, in cartesian and curvilinear orthogonal grid. The flow equations of motion are spatially discretized over a staggered grid and integrated using a finite-volume method, and advanced in time via three-step Runge-Kutta procedure, \textcolor{black}{where nonlinear terms are explicitly computed while the linear diffusive terms are treated using the semi-implicit Crank-Nicolson scheme.} Incompressibility is enforced at the end of the complete time step through a projection technique. Centered schemes are used to evaluate the spatial derivatives. The corresponding solution of the Navier–Stokes equations is second-order accurate in space and time on a uniform grid.\\

No-slip boundary condition is imposed at the channel walls, and periodic boundary conditions are applied for the velocity and perturbed pressure along the azimuthal (streamwise) direction. At the start of a simulation, the fluid is at rest. A pressure gradient is then imposed in the streamwise direction to induce the flow. The convergence to a fully steady and developed flow is obtained once the residuals of the velocity components tend towards small relative values (of the same order or less than $10^{-5}$). The evolution of the average flow velocity is recorded in time. The steady streamwise velocity, and subsequently the Reynolds numbers are determined \textit{a posteriori}, once the simulation reaches the steady state. \\

A uniform grid is used for the simulations: \textcolor{black}{the grid is rectangular in the radial and axial directions, and circular in the azimuthal direction.} The independence of the results on the mesh was ensured by testing four different grids described in table \ref{tab:tab_mesh} included in the appendix \ref{sec:mesh}, with $N_z$, $N_r$ and $N_\theta$ denoting the number of grid points along the channel height, width and  length, respectively. It should be noted that only a small section of the curved channel is considered in the numerical simulations, i.e., the total angle of the simulated domain is specified ($\Delta\theta = 5^\circ$), taking advantage of the periodicity in the azimuthal direction. Therefore, the mesh used in this work is significantly finer than in previous studies (for instance \cite{Fellouah2006}\cite{Chandratilleke2012}).\\

Most of the numerical simulations are carried out with $\lambda=\frac{3}{17}$ with a mesh resolution of $(N_r,N_\theta, N_)=(240,16,40)$, for five curvature ratios summarized in table \ref{tab:tab_delta}. The smallest curvature ratios $\delta_4$ and $\delta_5$ correspond to the curvature ratio $\delta$ of the smallest and largest loop of the microfluidic chips (Epigem) used for inertial focusing in \cite{howell2025experimental, hill2022efficient}, which consists of a 6-turn Archimedean spiral. Larger curvature ratios are interesting for experimental setups at larger scales \cite{Fellouah2006}\cite{humphrey1977}\cite{Baylis1971}. For each curvature ratio, simulations are carried out with different Reynolds numbers, in a way to cover Dean numbers between $1$ and $200$ approximately. \textcolor{black}{To assess the scaling of the radial velocity, an additional set of numerical simulations is carried out at a different aspect ratio, referred to as $\lambda_2=\frac{4}{17}$. Unless otherwise specified, the aspect ratio is $\frac{3}{17}$.}

\begin{table}[ht]
    \caption{\textcolor{black}{The five curvature radii ($mm$) and ratios considered in the simulations of the present work for a channel of $30 \times 170 \ \mu m$ (height x width)}.}
    \centering
    {
    \setlength{\tabcolsep}{18pt} 
    \begin{tabular}{|c|c|c|c|}
    \hline
        \textcolor{black}{$R_1$}  &   \textcolor{black}{1.85 }  & $\delta_1$    &    0.14  \\ \hline
        \textcolor{black}{$R_2$}  &   \textcolor{black}{2.85}   & $\delta_2$   &    0.089 \\ \hline
        \textcolor{black}{$R_3$}  &   \textcolor{black}{4.85 }  & $\delta_3$    &    0.053 \\ \hline
        \textcolor{black}{$R_4$}  &   \textcolor{black}{6.5  }  & $\delta_4$    &    0.039 \\ \hline
        \textcolor{black}{$R_5$}  &   \textcolor{black}{39.5 }  & $\delta_5$    &    0.006 \\ \hline
    \end{tabular}
    }
    \label{tab:tab_delta}
\end{table}

\noindent In the following sections, we present the effect of flow inertia and curvature on the primary and secondary flows, while varying the Dean number from small to moderate values. Directly related to the flow velocity is the friction factor which is also calculated.

\section{General flow features} \label{sec:primary_flow}

\begin{figure}[ht]
    \centering
    \includegraphics[scale=0.5]{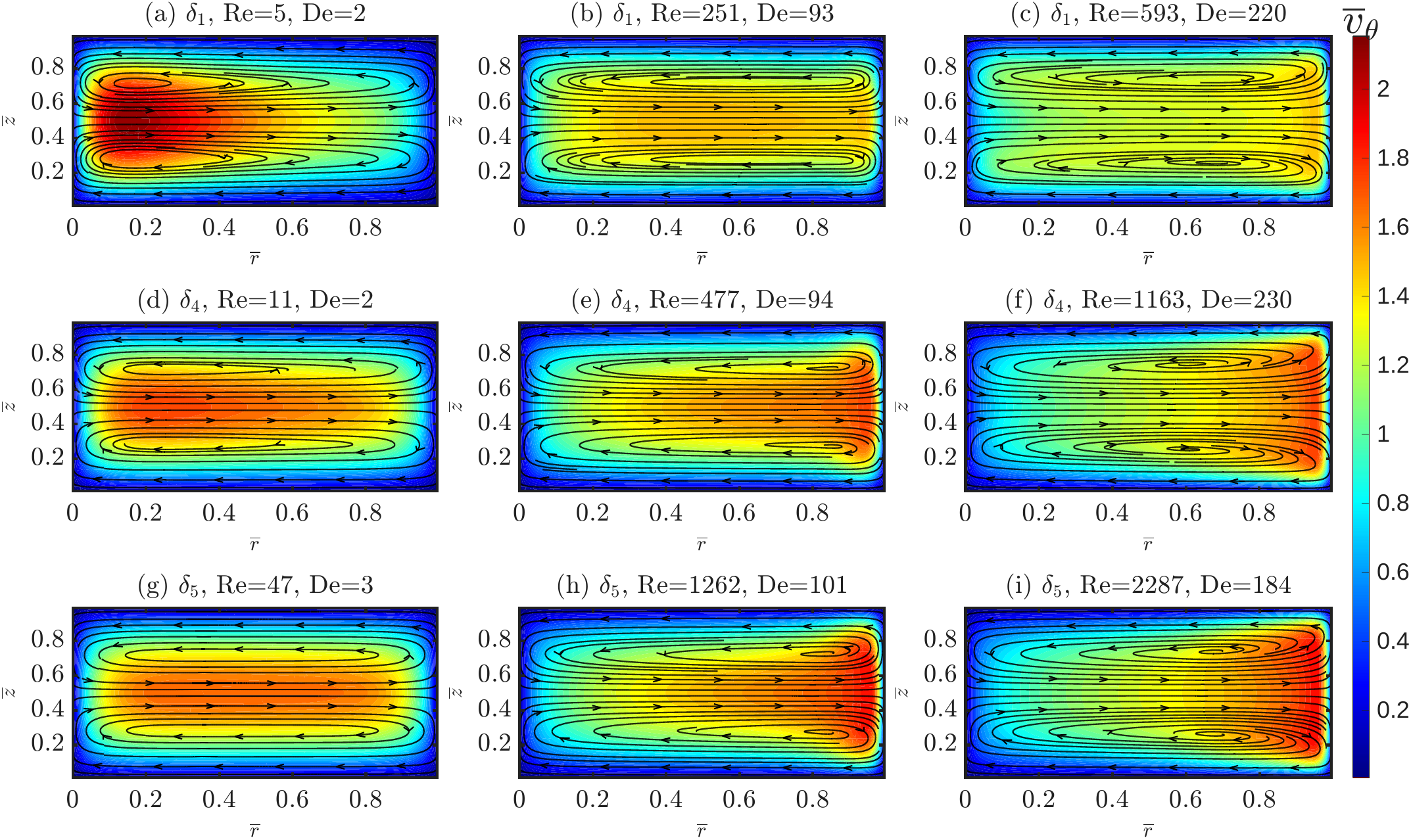}
    \caption{Contours of the azimuthal velocity $\overline{v}_\theta$ for various Dean numbers and curvature ratios. The streamlines indicate secondary vortices in the cross-section. Note that the aspect ratio of the channel cross-section is $\lambda=\frac{3}{17}$.} 
    \label{fig:mat3x3}
\end{figure}

\autoref{fig:mat3x3} displays the contours of the dimensionless azimuthal velocity $\overline{v}_\theta$  in the channel cross section, as well as the streamlines of the transverse velocities, once the steady state is reached. The rows of this figure correspond to curvature ratios $\delta_1, \delta_4 \text{ and } \delta_5$, respectively, and the columns correspond to $\mathrm{De}\sim 1$, $\sim 100$ and $\sim200$, respectively.

\noindent First let's examine the results at low Dean number. At high curvature (Fig.\ref{fig:mat3x3}a), the maximum azimuthal velocity $\overline{v}_\theta$ as well as the centers of the secondary vortices are located near the inner wall. At smaller curvature however, this maximum shifts to the middle of the channel (Fig.\ref{fig:mat3x3}g), similarly to the flow in a straight rectangular channel. As the Dean number increases, the secondary flow becomes stronger and both the peak of the streamwise velocity as well as the center of the secondary vortices shift towards the outer wall (Fig.\ref{fig:mat3x3} (c,f,i)).

Already at this stage of observation, our results confirm, what is known in curved channel or pipe flows \cite{berger1983}, that the secondary flows are not fully characterized by the Dean number and that other parameters such as the curvature ratio and the aspect ratio are key parameters to fully characterize the flow structure. In the following, we will add few comments on the main and secondary flow features. 

It is not common to find the peak of the streamwise velocity closer to the inner wall of a curved pipe or channel flow, since this occurs only at weak inertia. In the case where $\lambda\rightarrow\infty$, the parallel flow solution (with no secondary flow) is a shifted parabola toward the inner wall, sometimes referred to as CCPF (curved channel Poiseuille flow) \cite{bottaro1993}, as long as inertial effects are weak. While the velocity peak shifts from the inner wall to the outer wall as the Dean number is increased at large curvature, the pressure gradient in the radial direction remains positive for all curvature ratios and Dean numbers. The pressure peak falls at the outer wall independent of the operating conditions (see Appendix \ref{sec:gradP}). \\

At the highest Dean numbers, $De \sim 200$, we do not observe additional pair of secondary vortices. The experiments of Sugiyama \emph{et al.} \cite{sugiyama1983} suggested that the critical Dean number, above which two pairs of secondary vortices occur instead of one pair, is $De_c \sim 100 $ for $\lambda = 0.5$. They also suggested that $De_c$ increases when the channel aspect ratio decreases.
\textcolor{black}{Thus, the fact that only one pair of vortices is observed in the thin cross-section at moderate Dean numbers (in this work) agrees with their finding.} 
Recently, the work of Kim \emph{et al.} \cite{Kim2023} based on numerical simulations showed that the critical Dean number, above which two pairs of secondary vortices are observed, obeys the following empirical law: $\lambda^{1/2} \mathrm{De}_c \sim 110$. Our simulations agree with their finding, as the product $\lambda^{1/2} \mathrm{De}$ is smaller than 110, at the largest Dean numbers considered. This finding contradicts with the work of Nivedita \emph{et al.} \cite{Nivedita2017}, who observed based on fluorescence (with relatively low resolution) that two pairs of secondary vortices occur at lower Dean numbers  $De_c \sim 40 $. It is not possible to comment further on this disagreement, as to the authors' knowledge, there is no other work that show such transition at relatively low Dean number. \\

At large curvature ratio and for sufficiently large Dean numbers, transient structures take place in the flow (see Appendix \ref{sec:unstableFlow} for $\delta_1$ and $De \sim 230$) once the flow  has traveled a long distance, i.e. travel angle larger than $\approx 10\pi$ which corresponds to several turns if the channel was infinitely long with a small variation of the radius of curvature. One example of the flow structure is further discussed in appendix \ref{sec:unstableFlow}. We did not investigate thoroughly the flow transition, as it is out of the scope of the present work, but we expect that the transition angle $\theta_t$, when transition occurs, should depend on both the curvature ratio and the Reynolds (or Dean) number. For instance for $\delta_1$, these structures have appeared at shorter time in comparison with smaller $\delta$.

Next, we show quantitative features of the primary and secondary flow, based on velocity profiles. Furthermore, some velocity gradient profiles are also included in section \ref{sec:mesh}, in the frame of the numerical tool validation. 

\section{Velocity profiles } \label{sec:secondary_flow}

\begin{figure}[ht]
    \centering
    \includegraphics[width=0.84\linewidth]{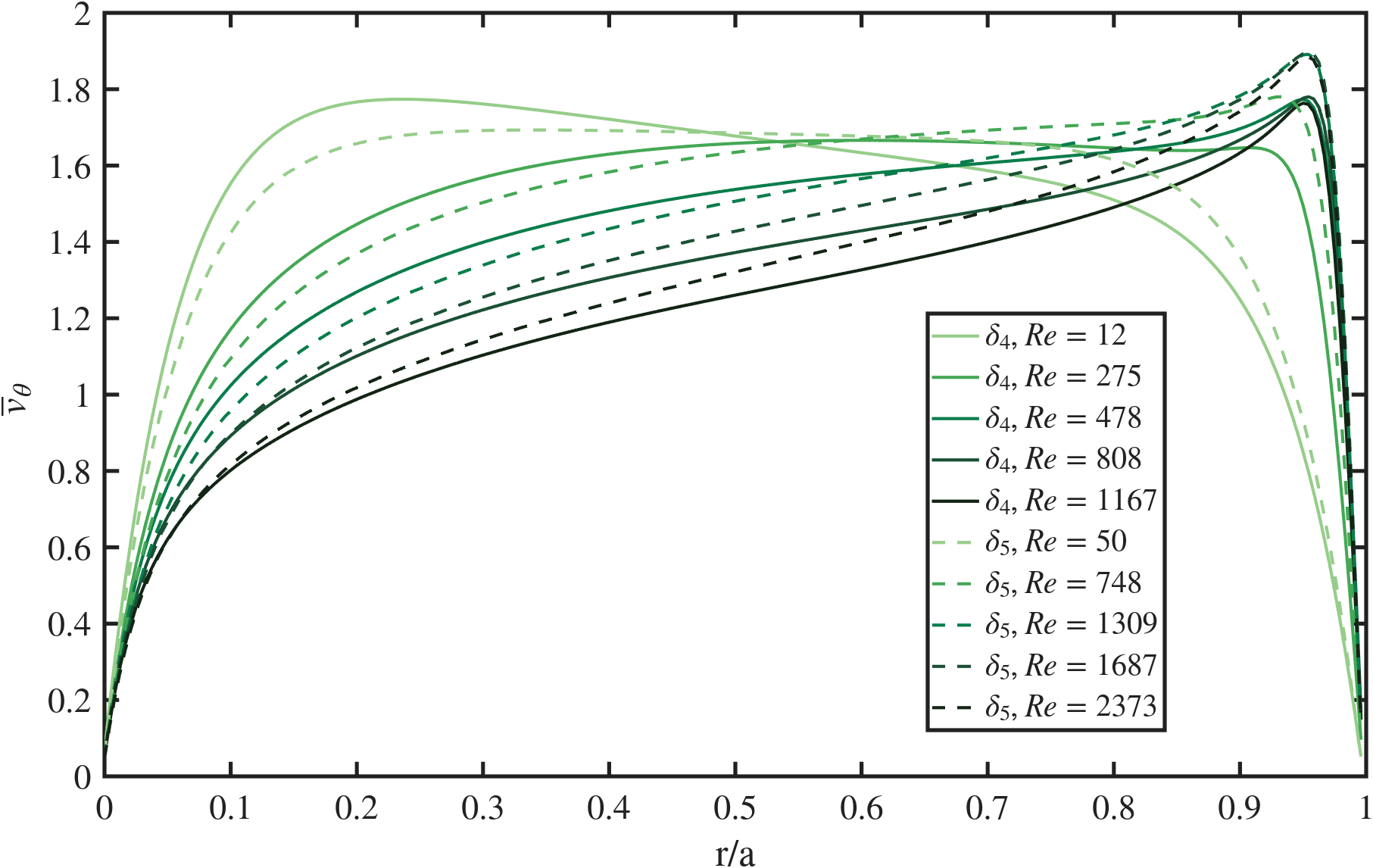}
    \caption{Profiles of the dimensionless tangential velocity $\overline{v}_\theta$  along the mid-plane $\overline{z} = 0.5$ for various $\mathrm{Re}$ and curvature ratios. Solid and dashed lines correspond to \textcolor{black}{$\delta_4\approx 0.039$ and  $\delta_5\approx 0.006$}, respectively.}
    \label{fig:Wscaled_d4_d5}
\end{figure}

\autoref{fig:Wscaled_d4_d5} displays the radial profile of the dimensionless azimuthal velocity along the mid-plane $\overline{z} = 0.5$ in the channels of curvature ratio $\delta_4$ and $\delta_5$ for various Reynolds numbers. Those profiles evidence the existence of a local velocity maximum near the inner wall of the channel of curvature ratio $\delta_4$ and the shift of the velocity maximum toward to the outer wall when the Reynolds number increases. In addition, we note that at high Reynolds (or Dean) numbers, the maximum is located at \textcolor{black}{a distance $\approx b/4$ from the outer wall (similar observation was noted with a channel of different aspect ratio $\lambda_2=\frac{4}{17}$).} The same conclusions can be drawn for $\delta_5$, although the velocity profiles at low De are flatter than those reported for $\delta_4$. 

\begin{figure}[!ht]
    \centering
    \subfloat[]{%
        \includegraphics[width=0.495\textwidth]{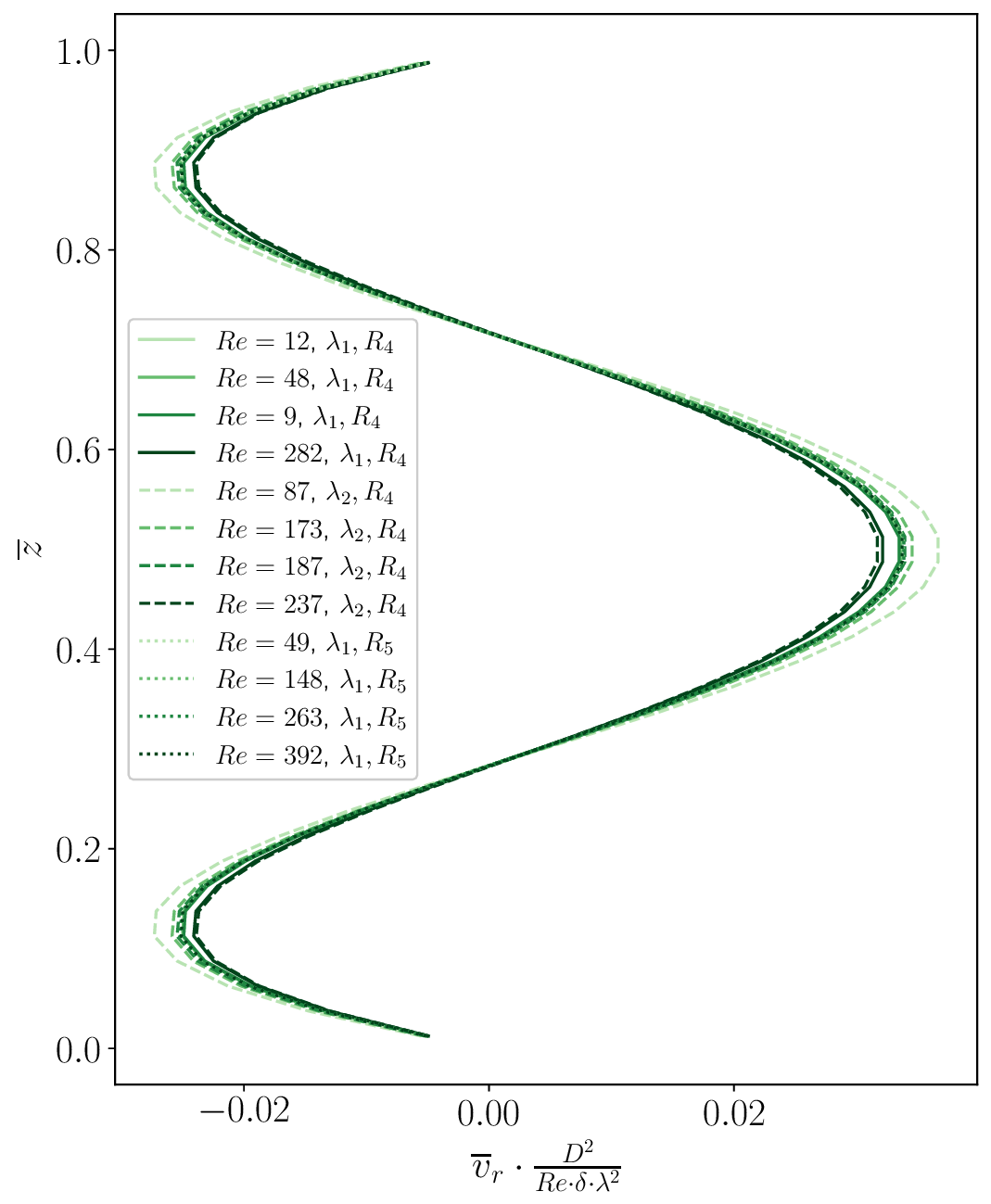}%
        \label{fig:Vr_low_De}}
    \hfill
    \subfloat[]{
        \includegraphics[width=0.495\textwidth]{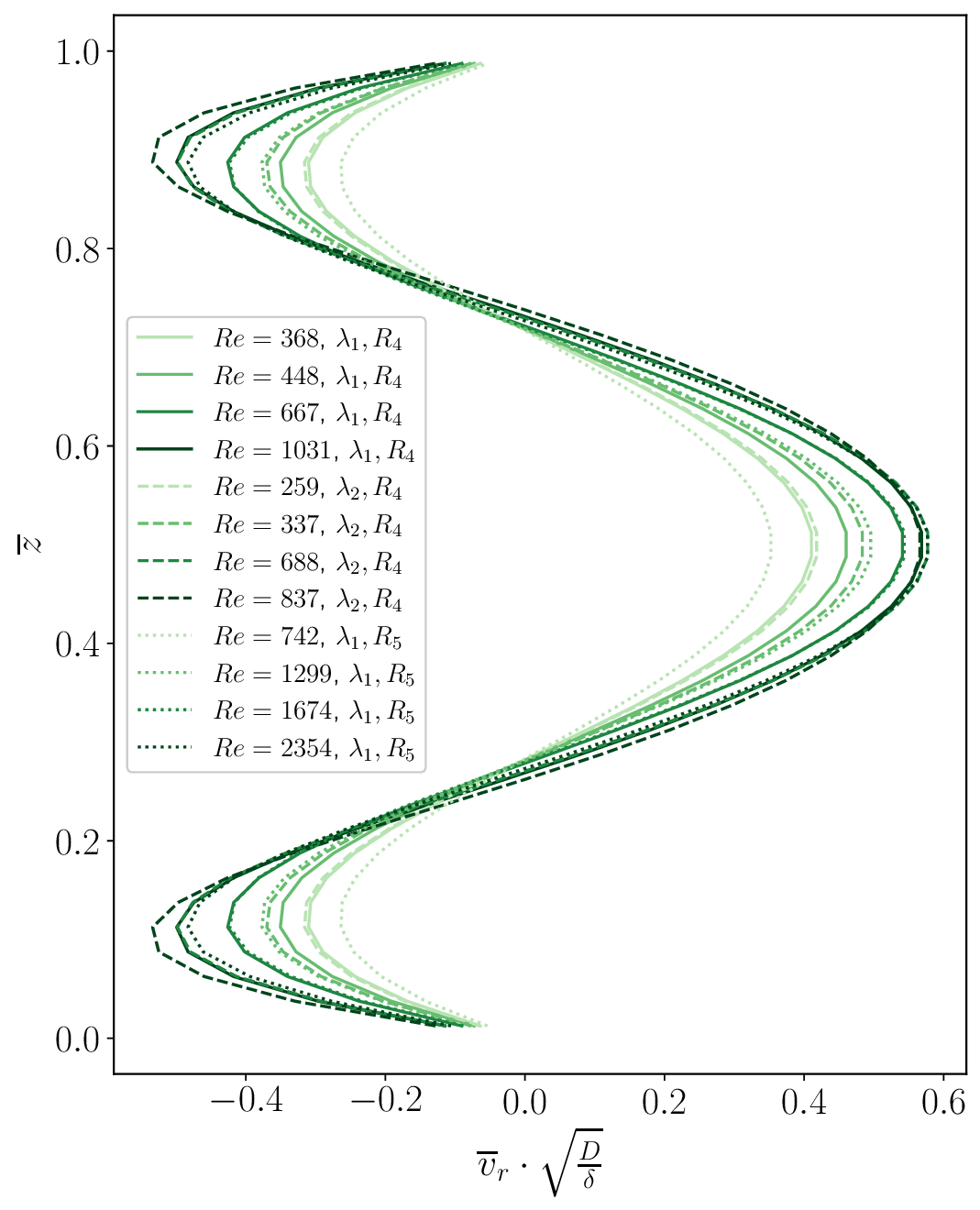}%
        \label{fig:Vr_mod_De}}
    \caption{Axial profiles of the radial velocity $\overline{v}_r$ along the mid-plane $\overline{r} =0$ for \textcolor{black}{two distinct aspect ratios ($\lambda_1=\frac{3}{17}, \lambda_2=\frac{4}{17}$) and two curvature radii ($R_4 , R_5$)} at (a) low De and (b) moderate De.}
    \label{fig:Vr_low_mod}
\end{figure}

\noindent Next, to better characterize the secondary flow, 
we consider the profile of the radial velocity $\overline{v}_r$, for instance along the axial direction, as displayed in Fig. \ref{fig:Vr_low_De} at the medium plane $\overline{r}=0$. As expected from the shape of secondary vortices, the radial velocity is positive near the midplane ($\overline{z}=0$), as it is oriented from the inner to the outer wall, and it is negative near the top and bottom walls. This figure shows also that scaling the radial velocity with \textcolor{black}{$\mathrm{Re} \ \delta  \ \frac{\lambda^2}{D^2}$ } at small Dean number $(\mathrm{De} \le 50)$, allows to collapse all the profiles. Moreover, at moderate Dean numbers ($50\le\mathrm{De}\le 250$), Fig. \ref{fig:Vr_mod_De} shows that the radial velocity scales reasonably well like \textcolor{black}{$\sqrt{\frac{\delta}{D}}$} which also aligns with the dimensional analysis presented in section \ref{sec:Dimensional}. The fact that the curves do not collapse on a single master trend suggests that $\overline{v}_r$ should also depend on the Reynolds number in addition to $\delta$. This will be discussed in the next section, considering the average amplitude of the secondary flow intensity. 

\section{Intensity of secondary flows}

In the cross-section of a curved channel flow, the velocity field takes the form of secondary vortices which intensity increases with the flow inertia.  It is important to understand how secondary flows scale with the dimensionless parameters, as they ensure mixing in the cross-section. 
At low Dean numbers, scaling in the form of power law have been suggested in the literature \cite{bayat2017semi}. At higher Dean numbers (up to 100), and using exact mathematical solution of the Navier-Stokes equations, Harding \cite{harding2022new} has shown that a simple power law scaling does not hold in the entire range of Dean numbers, suggesting another scaling, in the form of the inverse of a polynomial function.
Using numerical simulations, we will show that, following the dimensional analysis, one can find the correct scaling of the secondary flow in different regimes,  dominated by either viscous or inertial effects. \\

\begin{figure}[ht]
    \centering
    \subfloat[]{%
        \includegraphics[width=0.48\textwidth]{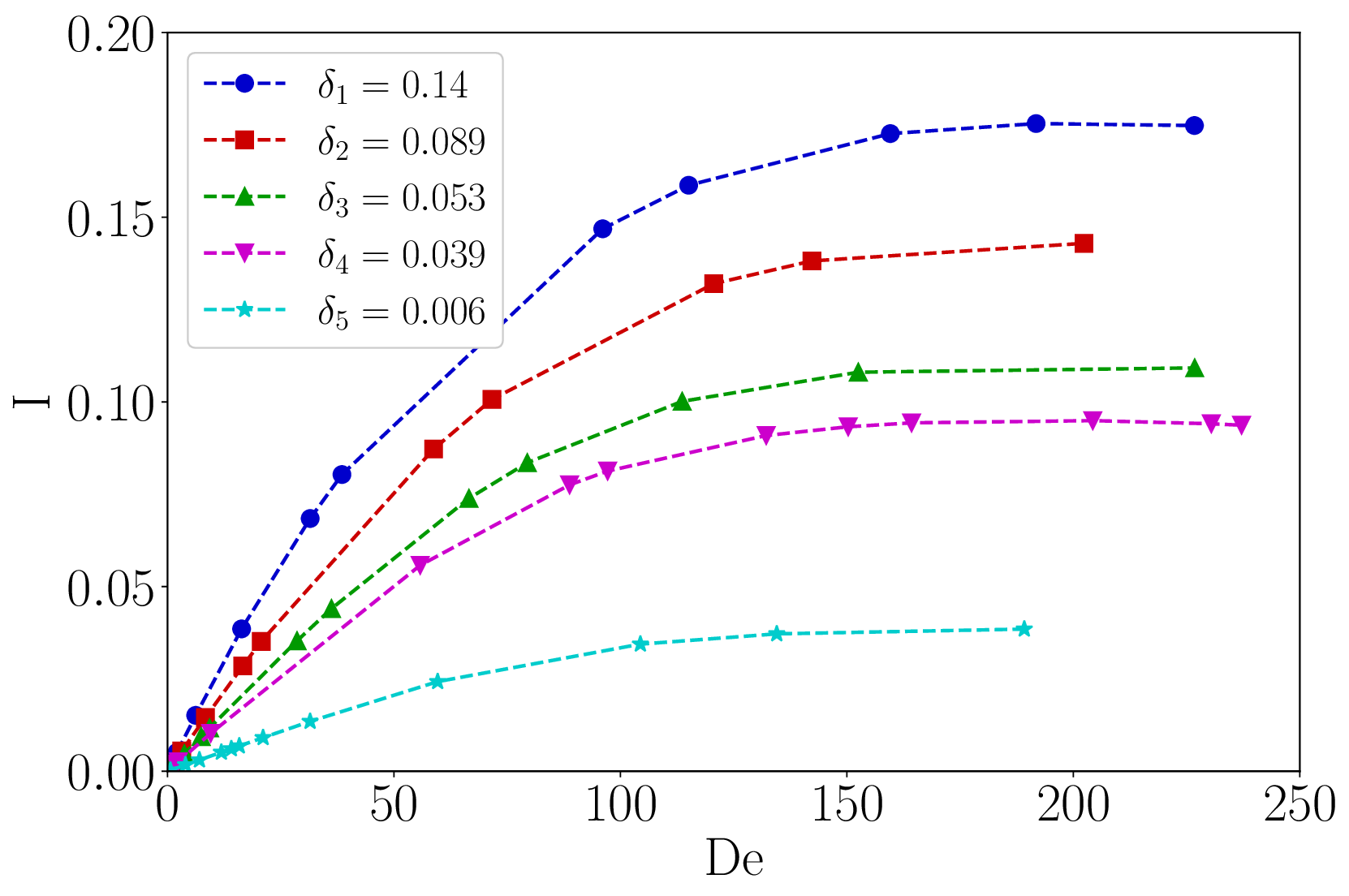}
        \label{fig:secondI}
    }
    \hfill
    \subfloat[]{%
        \includegraphics[width=0.48\textwidth]{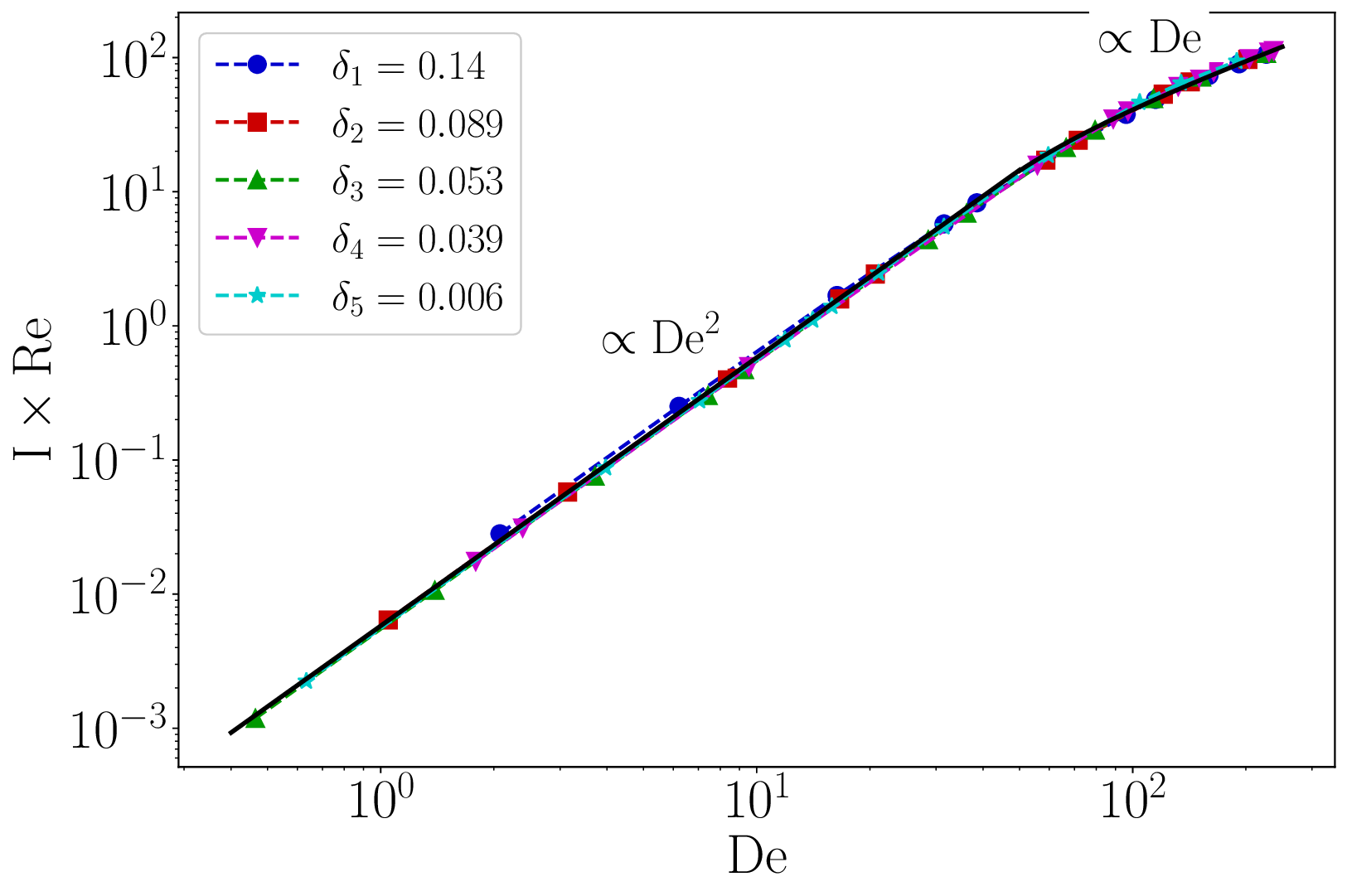}
        \label{fig:secondIxRe}}
    \caption{Secondary flow intensity as a function of the Dean number.}
    \label{fig:secondary_flow_intensity}
\end{figure}

The intensity of the secondary flow is defined from the root mean square of the radial and axial velocities:
\[
I= \dfrac{\left<\sqrt{v_r^2+v_z^2}\right>}{U_0} \equiv \left<\sqrt{\overline{v}_r^2+\overline{v}_z^2}\right>, 
\]

\noindent where the brackets denote the average over the cross-section.
Fig. \ref{fig:secondI} shows the secondary flow intensity $I$ as a function of the Dean number for different curvature ratios. At low to moderate Dean numbers, the intensity of secondary flows increases with the Dean number. As $I$ increases progressively starting from zero, this suggests the absence of bifurcation at low Dean numbers. The curves tend to level off starting from $\mathrm{De}\sim 100$. The plateau depends on the channel curvature: the stronger the curvature, the higher is the asymptotic value of the intensity $I$. When $\delta$ increases from 0.006 to 0.14, the intensity of secondary flows increases from 4 to 18\% of the average flow velocity. 

\noindent \ref{fig:secondIxRe} shows that, when plotting $I\times \mathrm{Re}$ as a function of the Dean number $\mathrm{De}$, the curves corresponding to different curvatures collapse onto a master curve. This figure shows that there are clearly two trends when the Dean number is increased. At low Dean numbers, $I\times \mathrm{Re}$ scales like $\mathrm{De^2}$ or equivalently $I\propto\delta\mathrm{Re}$, whereas at moderate Dean numbers, $I\times \mathrm{Re}$ scales rather like $\mathrm{De}$ or in an equivalent way $I\propto \sqrt{\delta}$. Two scaling laws are suggested in eq. \ref{eq:I_corr}, and are added to Fig.\ref{fig:secondary_flow_intensity}b within the corresponding range :  

\begin{subequations} 
\begin{align}
I &= \dfrac{1}{\mathrm{Re}} \left( 6\times10^{-3} \mathrm{De}^2 \right)  \text{ for }  De < 50  \\
I &= \dfrac{1}{\mathrm{Re}} \left(0.53\mathrm{De}-10.6 \right)  \text{ for }  50 < De < 250
\end{align}
\label{eq:I_corr}
\end{subequations}

\noindent \textcolor{black}{with a correlation factor $r^2$ equal to 0.9984 and 0.9953 for eqs. (\ref{eq:I_corr}a) and (\ref{eq:I_corr}b), respectively.}  The trends identified here are in agreement with the scaling analysis discussed in \autoref{sec:Dimensional} {\color{black} for a fixed aspect ratio $\lambda$}. 
While the contribution of the inertial terms to the momentum balance along the radial and axial directions (eqs. \ref{eq:NS_cyl}) is weak at low Dean numbers, the viscous and curvature terms balance each other, leading the dimensionless secondary velocity (components $\overline{v}_r$ and $\overline{v}_z$) to scale like $\delta\mathrm{Re}$. However, at moderate Dean numbers, the viscous terms have negligible contributions to the momentum balance, and thus the dimensionless secondary velocity is expected to scale like $\sqrt{\delta}$. It might seem surprising that the viscous scaling applies up to $De\sim 50$. We expect that this limit, as well as the constant appearing in the linear-like trend are dependent on the shape and dimension of the cross-section. \\

\section{Friction factor} \label{sec:friction_factor}

Although interesting from practical point of view (for dimensioning purposes), recent investigations on the fluid flow in curved channels do not systemically report the pressure drop.
This section aims thus to find a relationship between the pressure drop $\Delta P/L$ in its dimensionless form via the friction factor $f$, the Dean number and the curvature. The friction factor $f$ is defined as the ratio between the wall shear stress \(\tau_w =\dfrac{\Delta P }{L}\cdot \dfrac{d_h}{4}\) (this relation resulting from the macroscopic momentum balance), and the dynamic pressure, leading to $f =\dfrac{\tau_w}{\dfrac{1}{2} \rho  U_0^2}$. Replacing the wall shear stress by the pressure drop, the friction factor can be written : 

\begin{equation}
    f_c= \frac{1}{2R}\dfrac{\Delta P}{\Delta \theta} \cdot \dfrac{\rho d_h^3}{Re^2 \mu^2}
    \label{eq:fc}
\end{equation}

\noindent where the index $c$ refers explicitly to curved channels or pipes. The pressure drop in that case corresponds to $\dfrac{\Delta P }{L} = \dfrac{\Delta P}{R\Delta \theta}$, with \(\Delta\theta\) being the angle in radian across which the pressure drop is measured or imposed. Thus, for an imposed pressure drop $\dfrac{\Delta P}{\Delta \theta}$ in the simulations, the average streamwise velocity (or equivalently Reynolds number) is calculated once the steady state is reached, and subsequently, the friction factor $f_c$ is calculated following eq. \ref{eq:fc}. Then, the friction factor is scaled by the friction factor $f_s$ in a straight channel of similar cross section. The latter is calculated from numerical simulations carried out in a straight channel (with $\lambda=\frac{3}{17}$) and compared with the Darcy Weisbach relation valid for laminar flow \cite{white2011}: 
\[
f_s =\dfrac{k}{Re}
\]
where $k$ is a constant that depends on the shape and the aspect ratio of the channel cross-section (\cite{white2011}, p.387). For $\lambda=\frac{3}{17}$, the friction factor in the straight channel is $f_s =19/Re$.  \\

\begin{figure}[ht]
    \centering
    \includegraphics[width=0.85\linewidth]{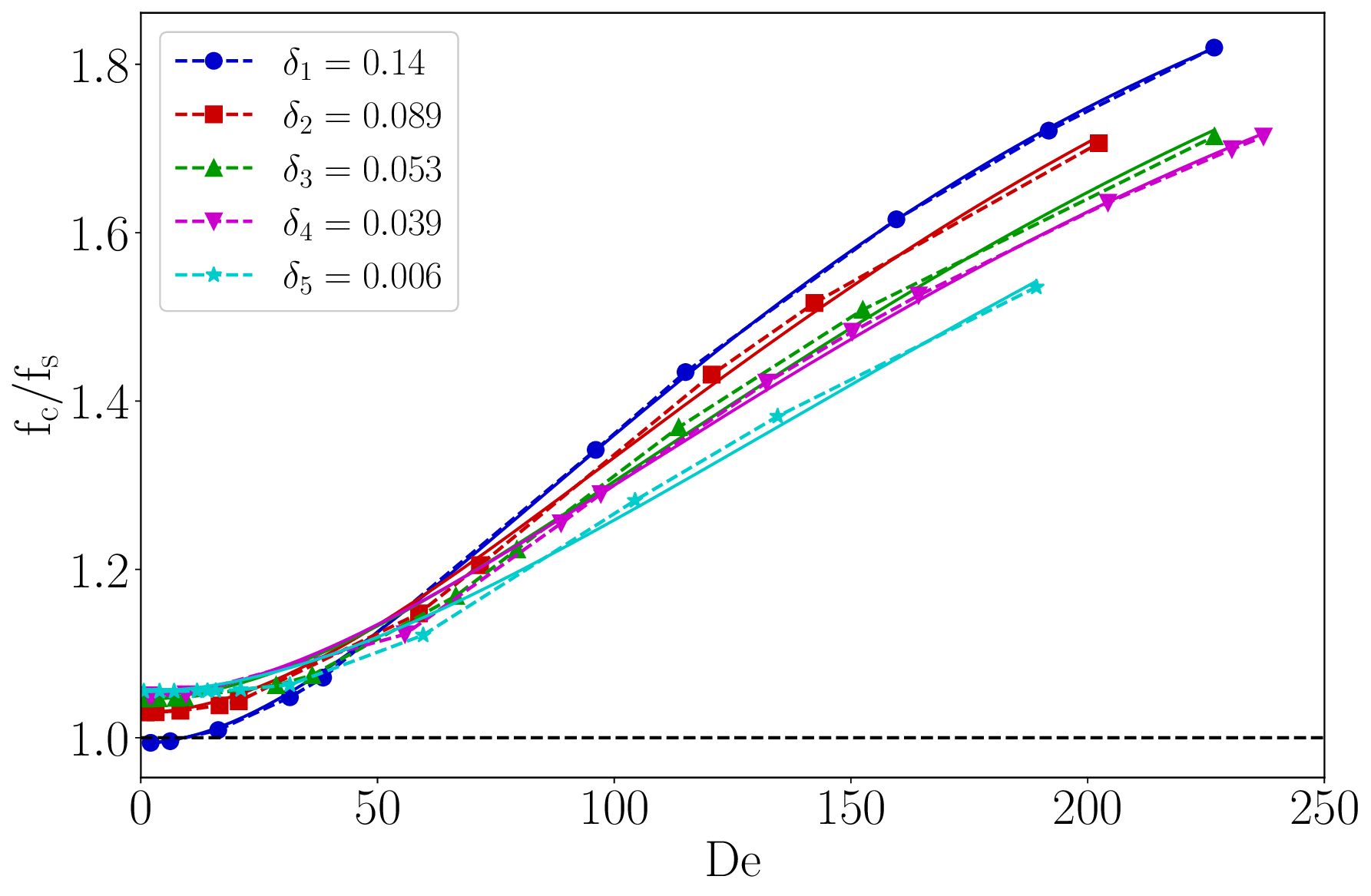}
    \caption{Friction factor of the curved channel flow, rescaled by the friction factor of a flow in a straight channel of similar cross-section, as a function of Dean number for different channel curvatures (indicated by symbols). Solid lines are from eq. \ref{eq:Cf_corr} and each color corresponds to one curvature ratio.}
    \label{fig:Cf_De}
\end{figure}

The rescaled friction factor $f_c/f_s$  is displayed in Fig. \ref{fig:Cf_De} as a function of the Dean number, for different curvature ratio. This figure shows that the rescaled friction factor increases with the Dean number. Our simulations suggest that the additional dependence of flow resistance on the curvature ratio $\delta$ is small beyond its dependence on the Dean number (which already includes the effect of curvature), in agreement with past studies in coiled pipes \cite{austin1973,truesdell1970numerical}. 
The plots at different $\delta$ follow this empirical law : 
\begin{equation}
    \dfrac{f_c}{f_s} = C + A \left[ 1- \left(1+\dfrac{\mathrm{De}}{\alpha} \right) \cdot \exp\left(\dfrac{-De}{\alpha}\right)\right]
    \label{eq:Cf_corr}
\end{equation}

\noindent where the parameters $C =\dfrac{f_c}{f_s} \bigg|_{\mathrm{De} \to 0}$, $A$ and $\alpha$ depend on the curvature ratio. \textcolor{black}{Those parameters, reported in table \ref{tab:tab_Cf_coeff}, are found such that eq. \ref{eq:Cf_corr} fits with numerical data with a correlation factor larger than 0.99 for all curvatures. }

\begin{table}[ht]
    \caption{Parameters of the normalized friction factor, eq. \ref{eq:Cf_corr}, for different curvature ratios. }
    \centering
    {
    \setlength{\tabcolsep}{20pt} 
    \begin{tabular}{|c|c|c|c|}
    \hline
                   & $C$ &  $A$ &  $\alpha$    \\ \hline
        $\delta_1$ & 0.99  &    1.1   &  84.44 \\ \hline
        $\delta_2$ & 1.03  &    1.12  &  98.86 \\ \hline
        $\delta_3$ & 1.04  &    1.09  &  108.12\\ \hline
        $\delta_4$ & 1.05  &    1.01  &  105.12\\ \hline
        $\delta_5$ & 1.055 &    1.18  &  133.78\\ \hline
    \end{tabular}
    }
    \label{tab:tab_Cf_coeff}
\end{table}

When $\mathrm{De}\rightarrow 0$, the rescaled friction factor $f_c/f_s\rightarrow C\approx 1$, and its dependence on the curvature ratio deserves some comments. $C$ is slightly smaller than 1, or equivalently the friction coefficient in the curved channel flow is slightly smaller than the friction factor in straight channel at the highest curvature, $\delta_1$. Similar observations were reported in the past, for instance in coiled pipes, where the flow resistance for a strongly curved pipe is less than that of a loosely coiled one \cite{lin1972laminar}. This effect can be attributed to the skewed streamwise velocity near the inner wall at large $\delta$ and small Dean number, as shown in the panel at the top left corner of Fig. \ref{fig:mat3x3}. Knowing that the surface of the inner wall is smaller than the surface of the outer wall in curved channels, the increase of the shear stress at the inner wall (compared to that on the wall of the straight channel) is thus compensated by the decrease of the inner wall surface, leading to a slightly smaller friction factor. On the contrary, $f_c/f_s$ tends to $\approx1.05$ instead of $1$ when the curvature is small. Several verification tests confirmed that this plateau is independent of the grid resolution in the three directions. At small $\delta$, the streamwise velocity is less skewed toward the inner channel wall, and since the surface of the outer wall is larger than the inner one, it leads to the slight increase of the friction factor compared to the straight channel flow. 

\section{Estimated entry length} \label{sec:entry_length}

Like the dimensionless pressure drop, the entry length is of practical importance as it gives insights on the distance required for the flow to become fully developed. 
An interesting picture of the flow development has been summarized by Berger et al. \cite{berger1983}. When a fluid is pumped in a pipe, the central core near the entrance is not influenced by the walls, unlike a thin layer near the wall that experiences viscous effects. The boundary layer develops initially like that in a straight pipe. Immediately downstream the entrance, the flow (in the streamwise direction) consists of two regions: the central region mainly experiences the centrifugal force (due to curvature) balanced by a radial pressure gradient, and the thin boundary layer where inertial and viscous forces balance each others. The inward pressure gradient induces a transverse flow in the boundary layer, in general from the outer wall toward the inner wall. The displacement effect in the boundary layer accelerates the flow in the core, while the secondary currents induce a flow from the inner to the outer wall. Those sequences occur up to a distance $R\theta_e=O(d_h)$ (or equivalently $\theta_e=O(\delta$) from the entry, for any Dean number. Beyond this distance, the flow depends on the Dean number. On the one hand, for small Dean number, the centrifugal force remains of second order (as in the first in the initial $O(\delta)$ region). The boundary layer continues to grow downstream until it fills the cross-section and the flow is fully developed. This length scale is $R\theta_e=O(d_h\mathrm{Re})$ like in a straight pipe, or equivalently $\theta_e=O(\delta\mathrm{Re})$. On the other hand, for large Dean numbers, centrifugal effects are as important as viscous and inertial effects, since the entrance. Much of the flow development occurs within a distance $R\theta_e=O(R\sqrt{\delta})\ll O(d_h\mathrm{Re})$, leading very short entry length in this case, compared to entry length in straight pipe.  \\

\begin{figure}[ht]
    \centering
    \subfloat[]{%
    \includegraphics[width=0.51\textwidth]{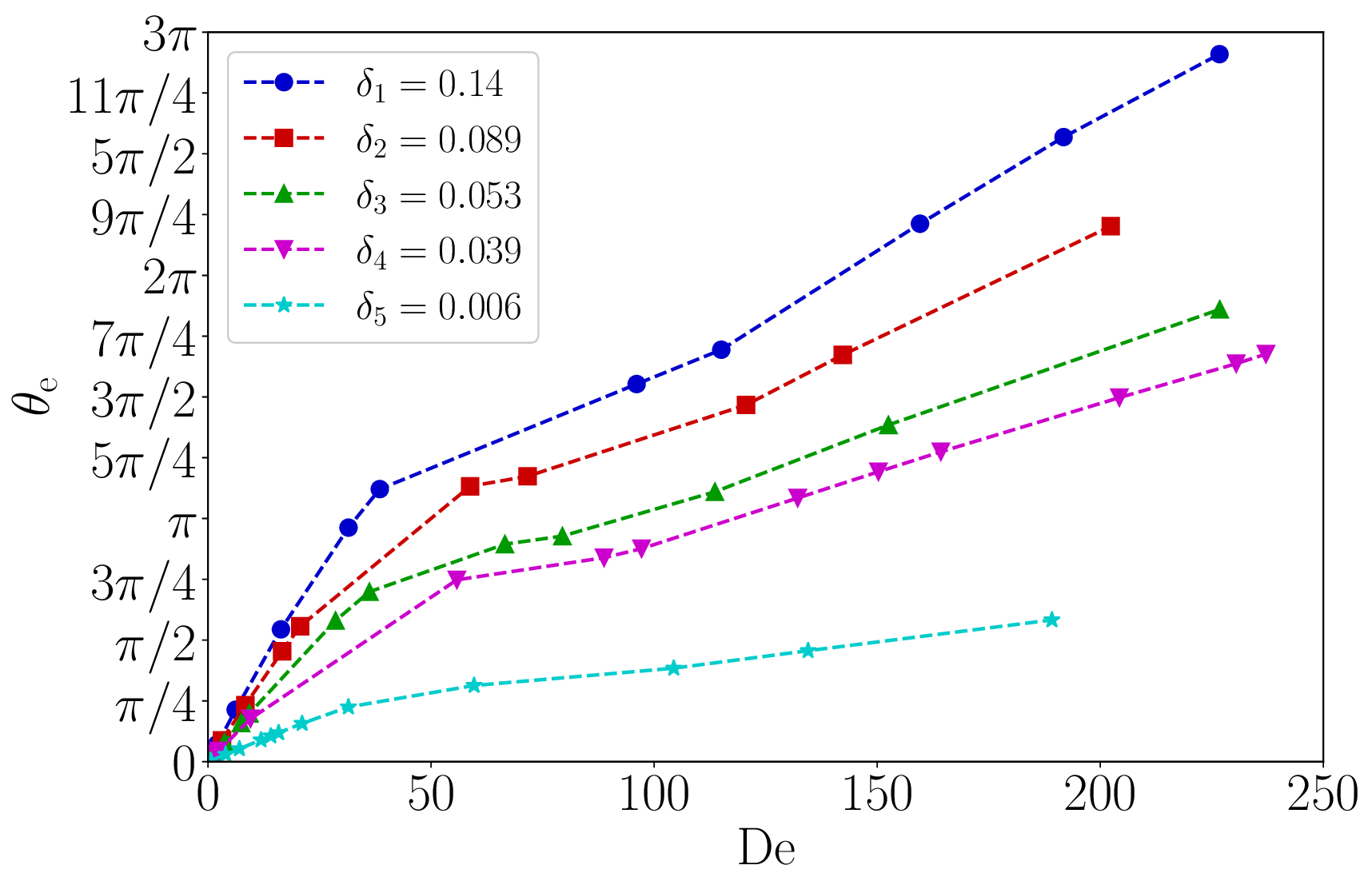}%
    \label{fig:theta_De}} 
    \hfill
    \subfloat[]{%
    \includegraphics[width=0.49\textwidth]{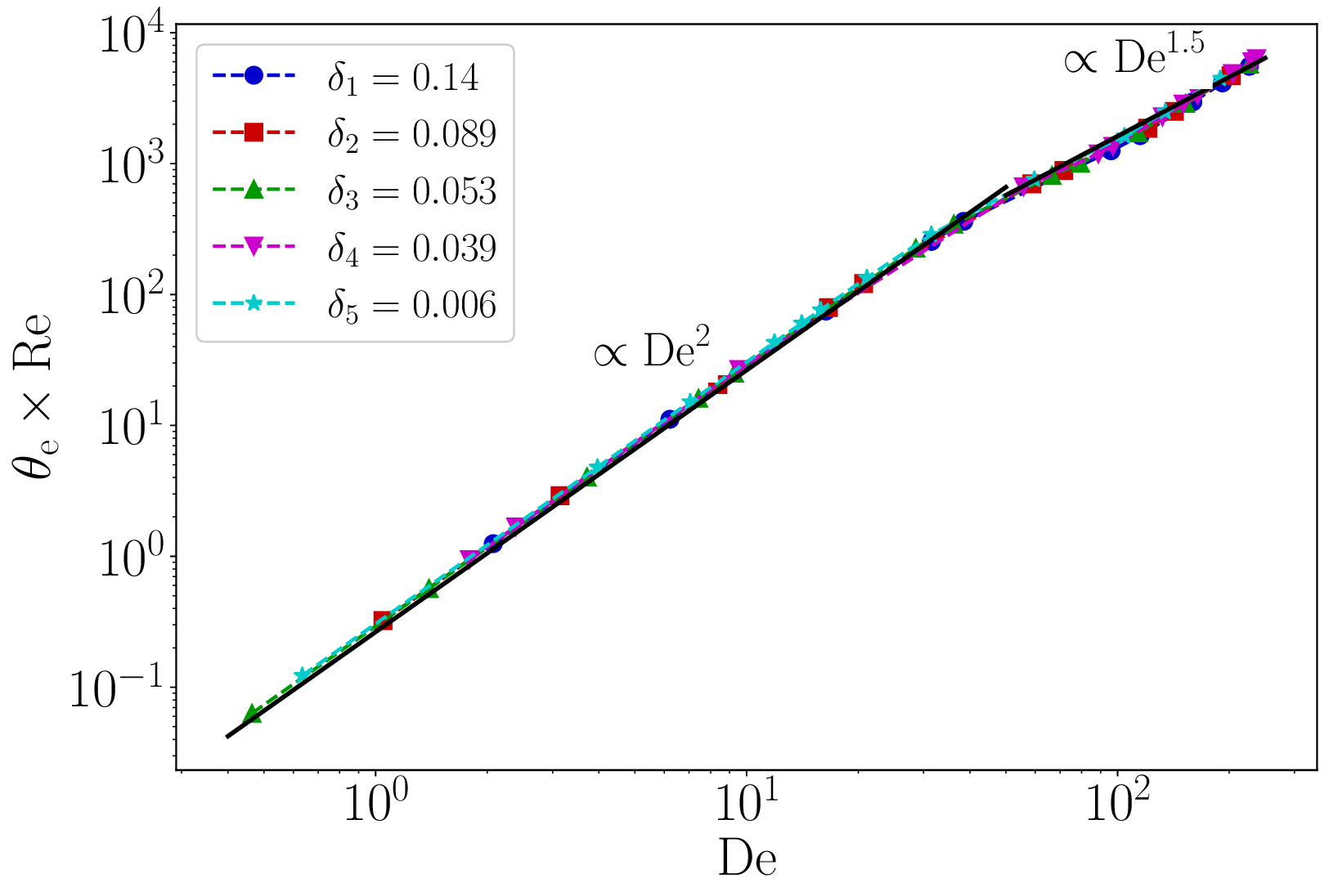}%
    \label{fig:theta_Redelta}}
    \caption{a) The entry angle $\theta_e$ as a function of the Dean number. In b) the entry angle is plotted in the form $\theta_e Re$ on logarithmic scale. The $\mathrm{De^2}$ and $\mathrm{De}^{1.5}$ scaling at small and moderate Dean numbers, respectively, are added to this panel. }
    \label{fig:theta}
\end{figure}

\textcolor{black}{The intention here is to obtain, from the numerical simulations, the scaling of the entry length or angle with the Reynolds number and the channel curvature. The flow evolving in time in periodic simulations mimics the flow development in space subsequent to flow entry at uniform velocity (such as the flow by connecting a pipe to a pump or to a large vessel with an abrupt change in cross-sectional area). Indeed, the distance traveled by the fluid during the transient stage is equivalent to the establishment length scale.}
From the numerical simulations, the entry length is calculated as following. The pressure drop is applied on the fluid initially at rest. During the first few time steps, the streamwise velocity rapidly evolves toward a quasi-uniform profile, driven by the volumetric forcing associated with the pressure gradient. Then, the fluid flow evolves progressively until reaching the steady state. 
To estimate the establishment length, we have calculate the instantaneous streamwise velocity averaged over the cross-section. The transient time scale $T_{L}$ is estimated from the time needed for the average streamwise velocity to reach \textcolor{black}{$99\%$} of its steady state value. The establishment length is thus calculated using
$R\theta_e=\int_{T_{L}}{\left<v\right>dt}$, with $\theta_e$ referring to the angle (in radian) traveled by the fluid, on average, before the fluid becomes fully developed. This "entry" angle $\theta_e$ is plotted as a function of the Dean number in Fig. \ref{fig:theta_De} for different curvature ratios. 
When multiplying $\theta_e$ by $\mathrm{Re}$, the curves collapse on a single global trend as illustrated in Fig. \ref{fig:theta_Redelta}. At small Dean numbers ($\mathrm{De} \le 50 $), $\theta_e Re$ scales like $De^2$ and therefore the entry angle scales like $\delta \mathrm{Re}$. 
\textcolor{black}{However, at moderate Dean numbers ($ 50 \le \mathrm{De} \le 250 $), we find that $\theta_e\mathrm{Re}$ scales like $\mathrm{De}^{1.5}$, which suggests that the entry angle scales like $\delta^{3/4} \mathrm{Re}^{1/2}$. }

\textcolor{black}{Thus, despite using an idealized configuration (periodic domain in the streamwise direction), the scaling compares well with the asymptotic theory in curved circular pipe flow both at low Dean number where centrifugal effects have weak impact on the boundary layer development \cite{berger1983,ault2017entry} and at moderate Dean number where centrifugal forces have significant effect \cite{yao1975entry}. Moreover, Fig. \ref{fig:theta_De} suggests that that the total angle (in radian) is smaller than $3\pi$ in all the simulations realized for this work. }

\section{Concluding remarks} \label{sec:conclusion}

This present work investigates the flow dynamics in curved channels with small aspect ratio, from small to moderate Dean numbers ($De \lesssim 250$) and for a wide range of curvature ratios ($0.005 \leq\delta\leq 0.15$). Most of the results are obtained with an aspect ratio $\lambda=\frac{3}{17}$. Streamwise velocity contours show a transition of the location of the velocity peak from near inner to near outer wall when the Dean number is increased and / or the curvature ratio is decreased. 
As for the secondary flow, only one pair of secondary vortices is found in the steady state regime.  
The dimensionless intensity of the secondary flows scale like $\delta \mathrm{Re}$ at small Dean numbers ($\mathrm{De} \le 50$), whereas they scale like $\sqrt{\delta}$ at moderate Dean numbers ($ 50 \le \mathrm{De} \le 250$), in agreement with the dimensional analysis. 
A law is suggested for the friction coefficient normalized by the friction coefficient in a straight duct of similar cross-section. 
As for the entry angle, it scales as $\delta \mathrm{Re}$ at small Dean number and as $\delta^{3/4} \mathrm{Re}^{1/2}$ at moderate Dean numbers, akin the entry angle of the flow in curved circular pipes. \\

The present set of results will enable the interpretation of particle transport in curved channel flow with small cross-section aspect ratio, commonly used for particle separation / concentration in the inertial regime. 
For instance, i) the alteration of the velocity peak along the radial direction can significantly impact hydrodynamic forces experienced by particles when transported by the curved flow, especially the radial lift near the inner wall, and ii) the scaling of the secondary flow intensity is useful to predict the drag force and the mixing time scale associated with the secondary flow. While our results stem from simulations of fluid flow induced by a pressure gradient, we expect that the conclusions also apply for flow in spiral channels with continuously increasing or decreasing radius of curvature, with small curvature variation along the flow direction, especially when the inertial effects are relatively weak, at low Dean numbers ($\mathrm{De} \le 50$), in agreement with the mathematical analysis of \cite{harding2018fluid}. At larger Dean numbers, the (in)dependence of the results on the curvature variation has yet to be discovered.

\section{Acknowledgment}
This research was carried as part of the project Posseidon-TIRIS co-funded by ANR under France 2030 program (ANR-22-EXES-0015), the Occitanie Region and the European Regional Development Fund. Computational resources were provided by the meso-centre CALMIP under project No.~P1002. The authors acknowledge support of EUR BioEco (grant ANR-18-EURE-0021). 

\section{Appendix} \label{sec:appendix}
\subsection{Accuracy of numerical solution} \label{sec:mesh}

We have carried out several tests in order to check the numerical accuracy of the code Jadim for solving curved channel flows. Table \ref{tab:tab_mesh} summarizes the tests in the case of a curvature ratio $\delta = 0.14$ and a rectangular cross-section with $\lambda=\frac{3}{17}$
and a total angle of 5° in the azimuthal direction. The Dean number is equal to 117 in that case. 

\begin{table}[ht]
    \caption{Distribution of the mesh in different directions.}
    \centering
    {
    \setlength{\tabcolsep}{10pt} 
    \begin{tabular}{c|c|c|c|c|c|c}
    \hline
               &  $N_z$  &  $N_r$   &   $N_\theta$   &    Number of  cells    &  $I$ & $\frac{f_c}{f_s}$ \\ \hline
        mesh 1 &  40  &  240  &   16   &    153600 &0.1629 & 1.476 \\ \hline
        mesh 2 &  40  &  360  &   16   &    460800 &0.1629 & 1.476 \\ \hline
        mesh 3 &  50  &  400  &   16   &    320000 &0.1628 & 1.458 \\ \hline
        mesh 4 &  50  &  480  &   16   &    768000 &0.1628 & 1.458 \\ \hline
    \end{tabular}
    }
    \label{tab:tab_mesh}
\end{table}

\begin{figure}[ht]
    \centering
    
    \subfloat[Non-dimensional tangential velocity $\overline{v}_{\theta}$ profile along the mid-plane $\overline{z} =0.5$]{
        \includegraphics[width=0.48\textwidth]{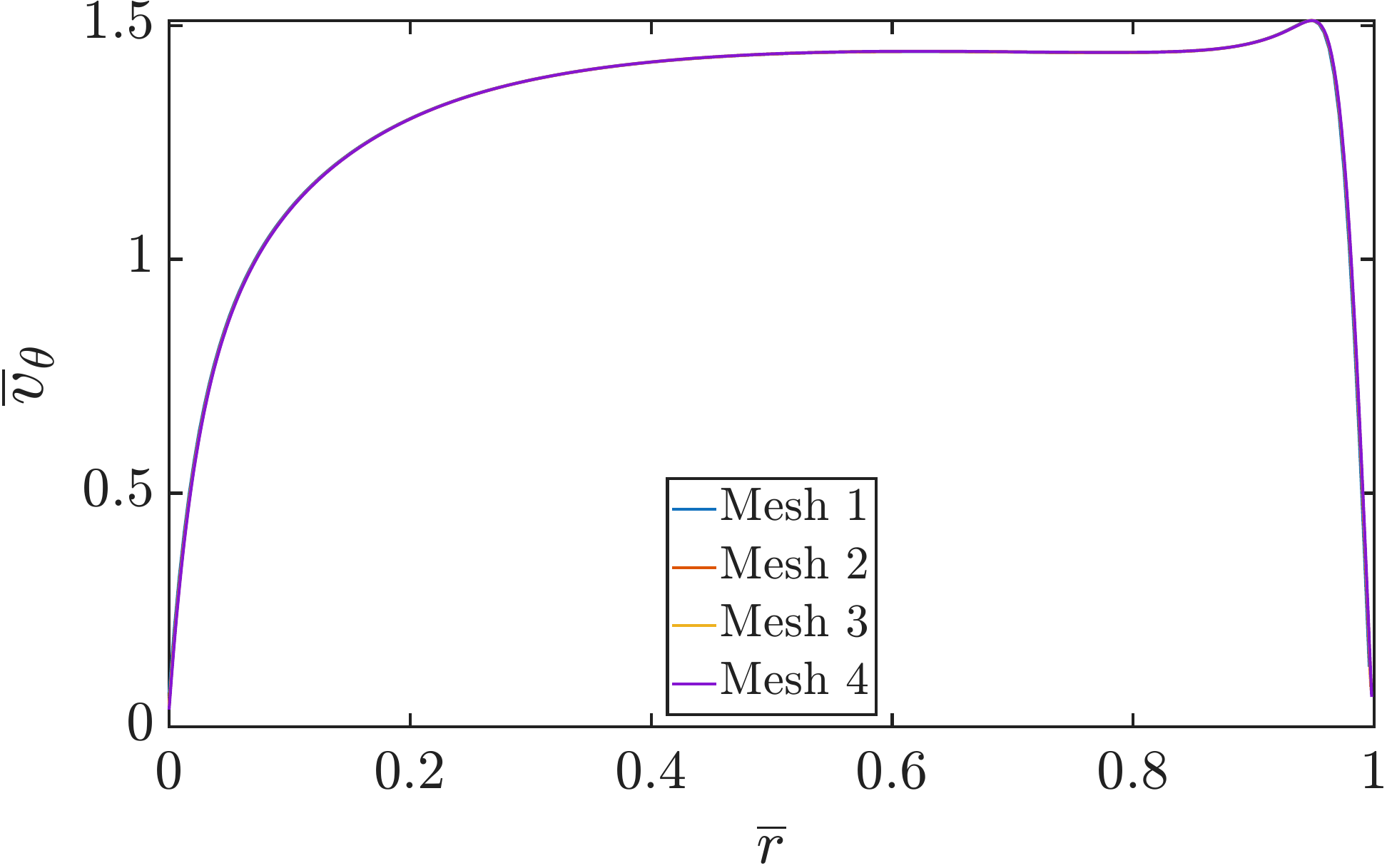}%
        \label{fig:mesh_subfig1} }
    \hfill
    \subfloat[Radial gradient of the tangential velocity $\partial \overline{v}_\theta /\partial  \overline{r}$ along the mid-plane  $\overline{z} =0.5$]{%
        \includegraphics[width=0.48\textwidth]{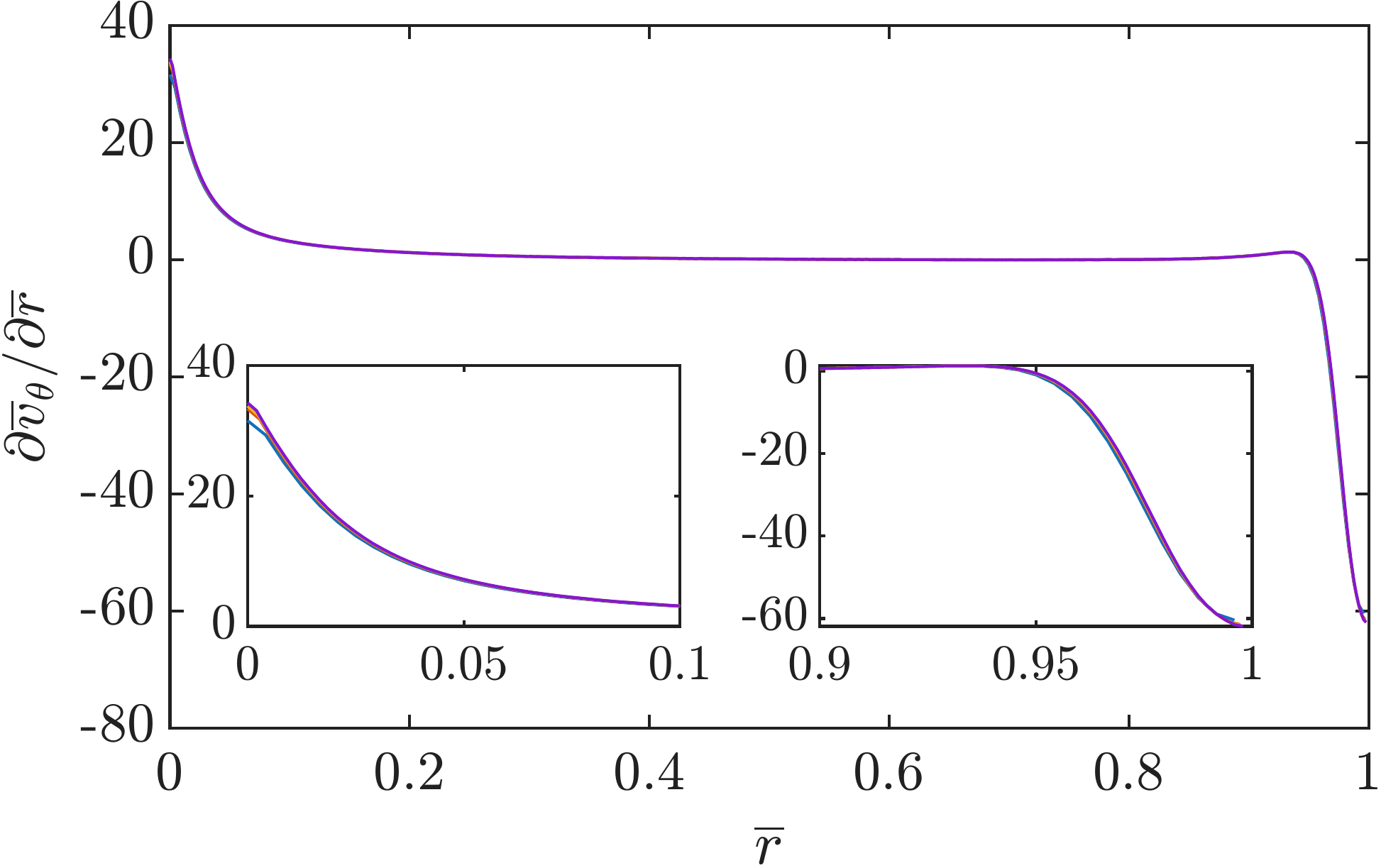}%
        \label{fig:mesh_subfig2} }
    
    \vspace{0.3cm}

    \subfloat[Axial gradient of the tangential velocity $\partial \overline{v}_\theta /\partial  \overline{z}$ along the mid-plane  $\overline{r} =0.5$]{%
        \includegraphics[width=0.49\textwidth]{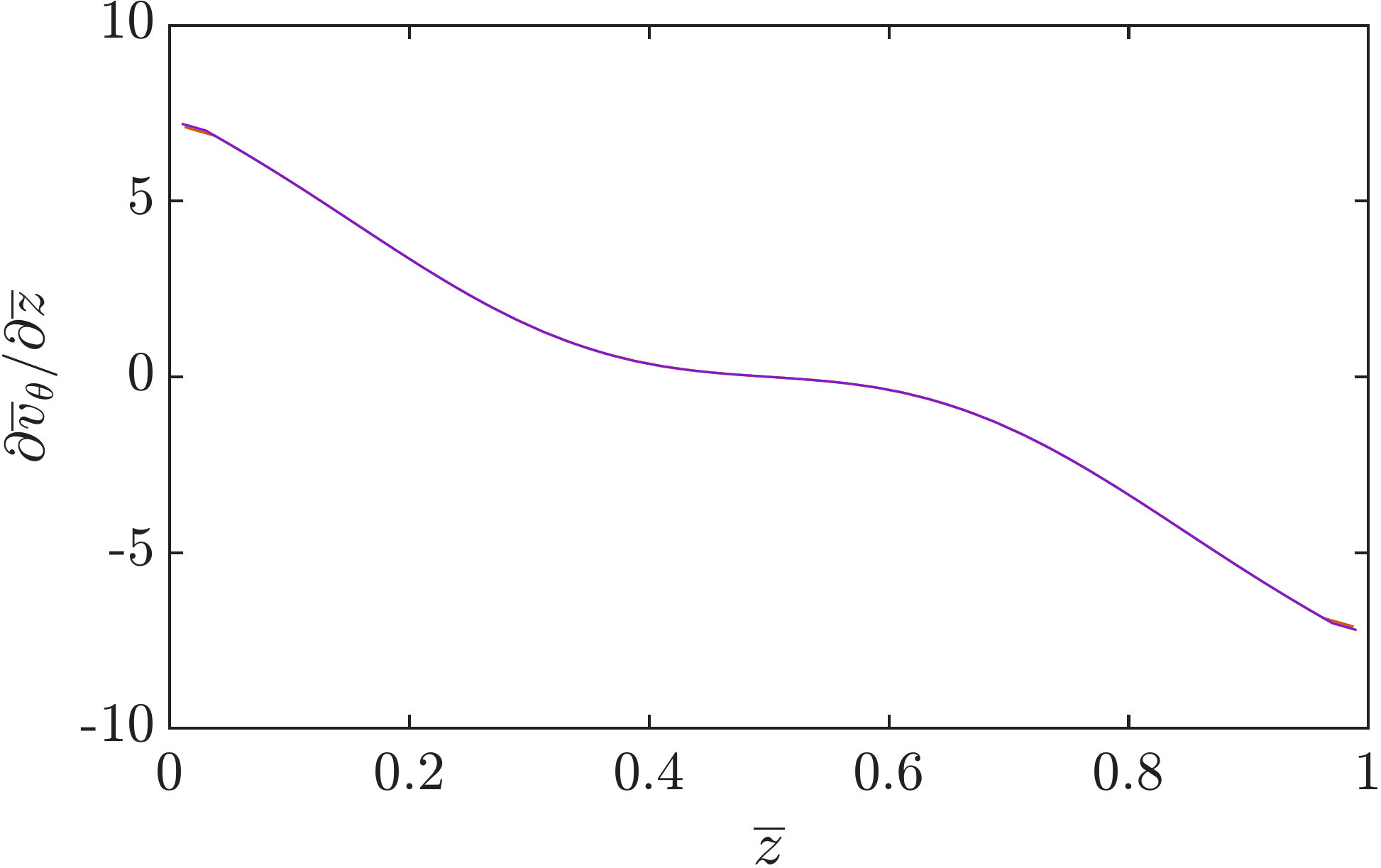}%
        \label{fig:mesh_subfig5}}
    \hfill
    \subfloat[Non-dimensional radial velocity $\overline{v}_r$ profile near the inner wall $\overline{r} =0.9$]{%
        \includegraphics[width=0.48\textwidth]{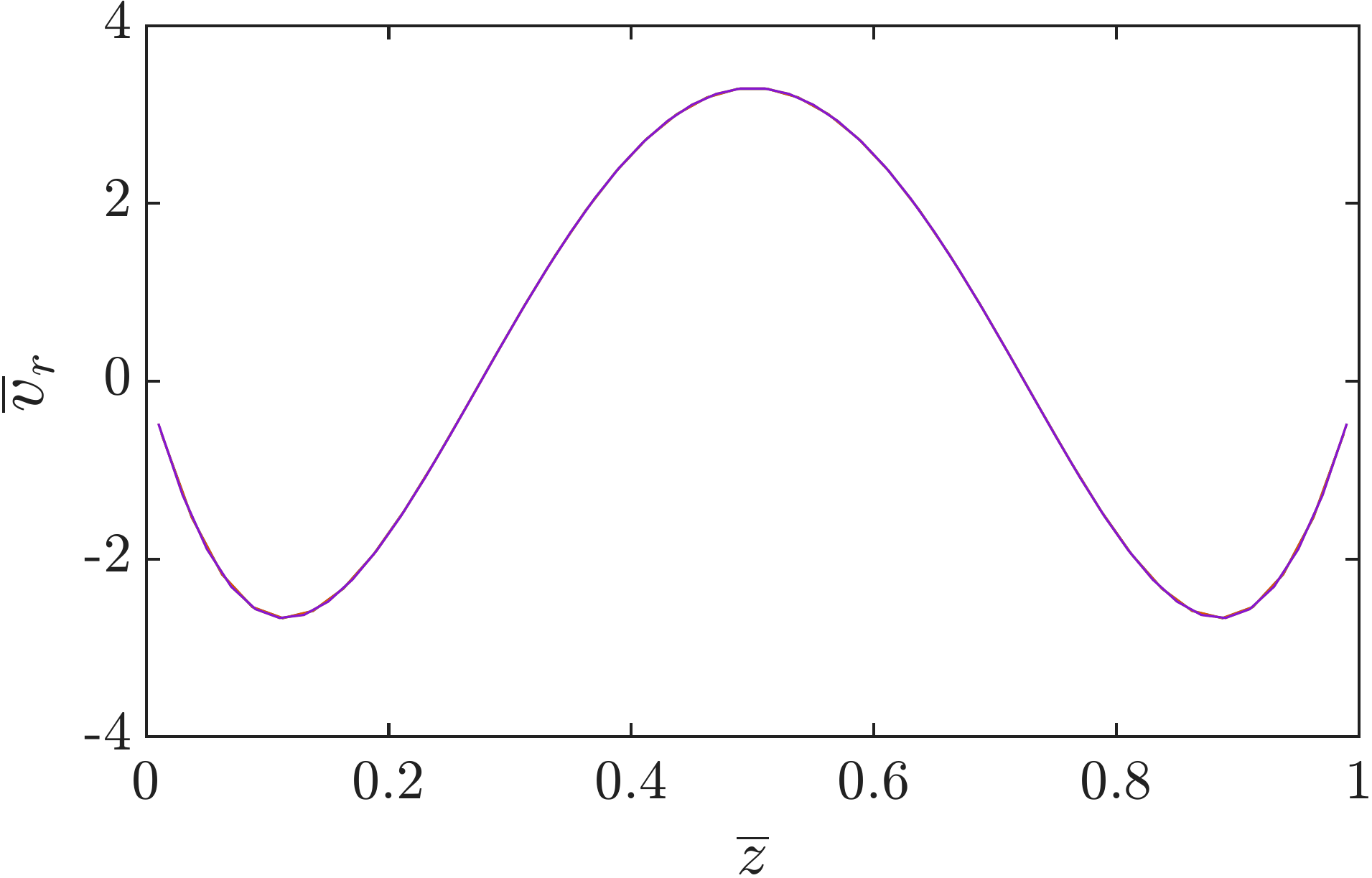}%
        \label{fig:mesh_subfig6}}
        
     \caption{Dependence on the grid resolution of the dimensionless profiles of the a) streamwise velocity, b) radial and c) axial gradient of the streamwise velocity, and d) radial velocity. The simulations were carried out in curved channel flow with $\delta_1$ and $\mathrm{De}=117$. The mesh distribution is given in table \ref{tab:tab_mesh}.}
\label{fig:mesh_dependence}
\end{figure}

Fig.~\ref{fig:mesh_dependence} illustrates the profiles of the dimensionless streamwise velocity $\overline{v}_\theta$ and its gradient along the radial and axial directions, as well as the radial velocity $\overline{v}_r$. Those profiles have been chosen since they involve the main quantities explored in this work, especially the velocity and friction factor. The profiles are almost overlapping.  \textcolor{black}{The maximum measured discrepancy corresponds to the difference in $\frac{\partial \overline{v}_\theta}{\partial \overline{r}}\mid_{\overline{r}=0}$ between simulations with different meshes is equal to $8\%$ of its value for the finer mesh. This mainly occurs on the side walls. The discrepancy of the velocity gradient at the top and bottom walls is negligible, and thus the effect of the mesh size on the friction coefficient was found to be negligible. } This suggests that the smallest grid resolution is already sufficient for studying rigorously the flow in the geometry of interest. 

Furthermore, the dependence of the results on the mesh length was tested by two ways. On the one hand, the curvature angle of the channel is changed as reported in table \ref{tab:mesh_angle}, while $N_\theta$ was kept fixed. The simulation results are independent of the domain length, , such as the friction coefficient and the intensity of secondary flows for reference. Alternatively, we kept the domain angle constant, i.e. $\Delta\theta=5^\circ$ and refined the mesh in the azimuthal direction by increasing $N_\theta$ from 16 to 64. This refinement likewise had no significant impact on the simulation results.

\begin{table}[ht]
    \caption{Mesh grids for various domaine angle  }
    \centering
    {
    \setlength{\tabcolsep}{10pt} 
    \begin{tabular}{c|c|c|c|c|c|c}
    \hline
               &  $N_z$  &  $N_r$   &   $N_\theta$   &    $\theta$    &  $I$ & $\frac{f_c}{f_s}$ \\ \hline
        mesh 1 &  40  &  240  &   16   & $5^\circ$  & 0.1629 & 1.4760 \\ \hline
        mesh 5 &  40  &  240  &   32   & $10^\circ$  & 0.1629 & 1.4760 \\ \hline
        mesh 6 &  50  &  400  &   64   & $20^\circ$  & 0.1629 & 1.4760 \\ \hline
    \end{tabular}
    }
    \label{tab:mesh_angle}
\end{table}

{\color{black}
Additionally, we fixed the domain angle (i.e. $\Delta\theta=5^\circ$) and refined the mesh along the azimuthal direction by increasing $N_\theta$ as reported in Table~\ref{tab:mesh_Ntheta}. Simulation results such as the friction coefficient and the intensity of secondary flows have not been significantly impacted.

\begin{table}[ht]
    \caption{Mesh grids for $\theta =5^\circ$ and various $N_\theta$ }
    \centering
    {
    \setlength{\tabcolsep}{10pt} 
    \begin{tabular}{|c|c|c|c|c|c|}
    \hline
               &  $N_z$  &  $N_r$   &   $N_\theta$    &  $I$ & $\frac{f_c}{f_s}$ \\ \hline
        mesh 1 &  40  &  240  &   16    & 0,16291 & 1.4760 \\ \hline
        mesh 7 &  40  &  240  &   32   & 0,16291 & 1.4760 \\ \hline
        mesh 8 &  40  &  240  &   64    & 0,16293 & 1,4764 \\ \hline
    \end{tabular}
    }
    \label{tab:mesh_Ntheta}
\end{table}
}

\subsection{Radial pressure gradient}\label{sec:gradP}

\begin{figure}[ht]
    \centering
    \subfloat[]{%
        \includegraphics[width=0.85\textwidth]{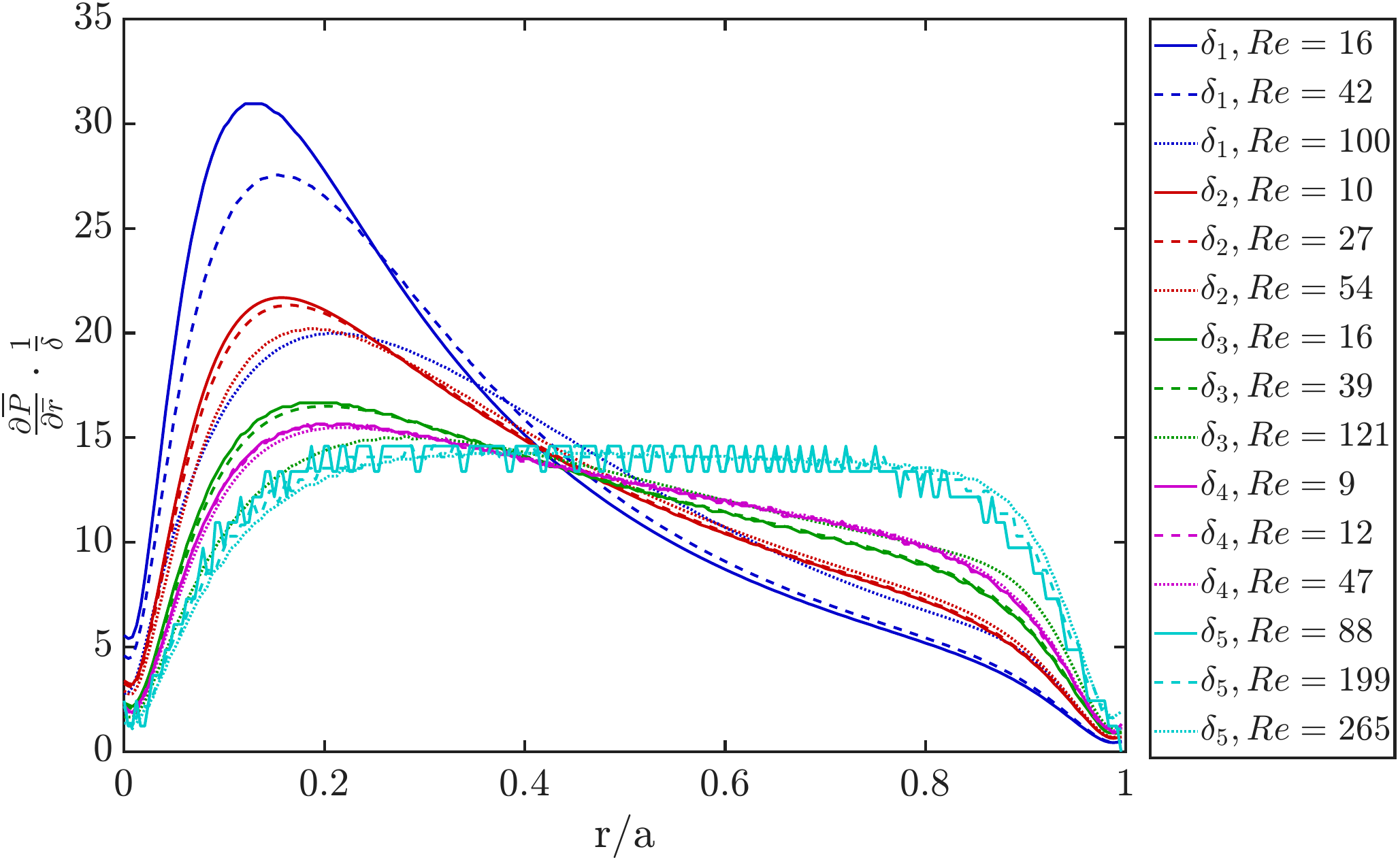}%
        \label{fig:dPdr_low_De}}
    \hfill
    \subfloat[]{%
        \includegraphics[width=0.85\textwidth]{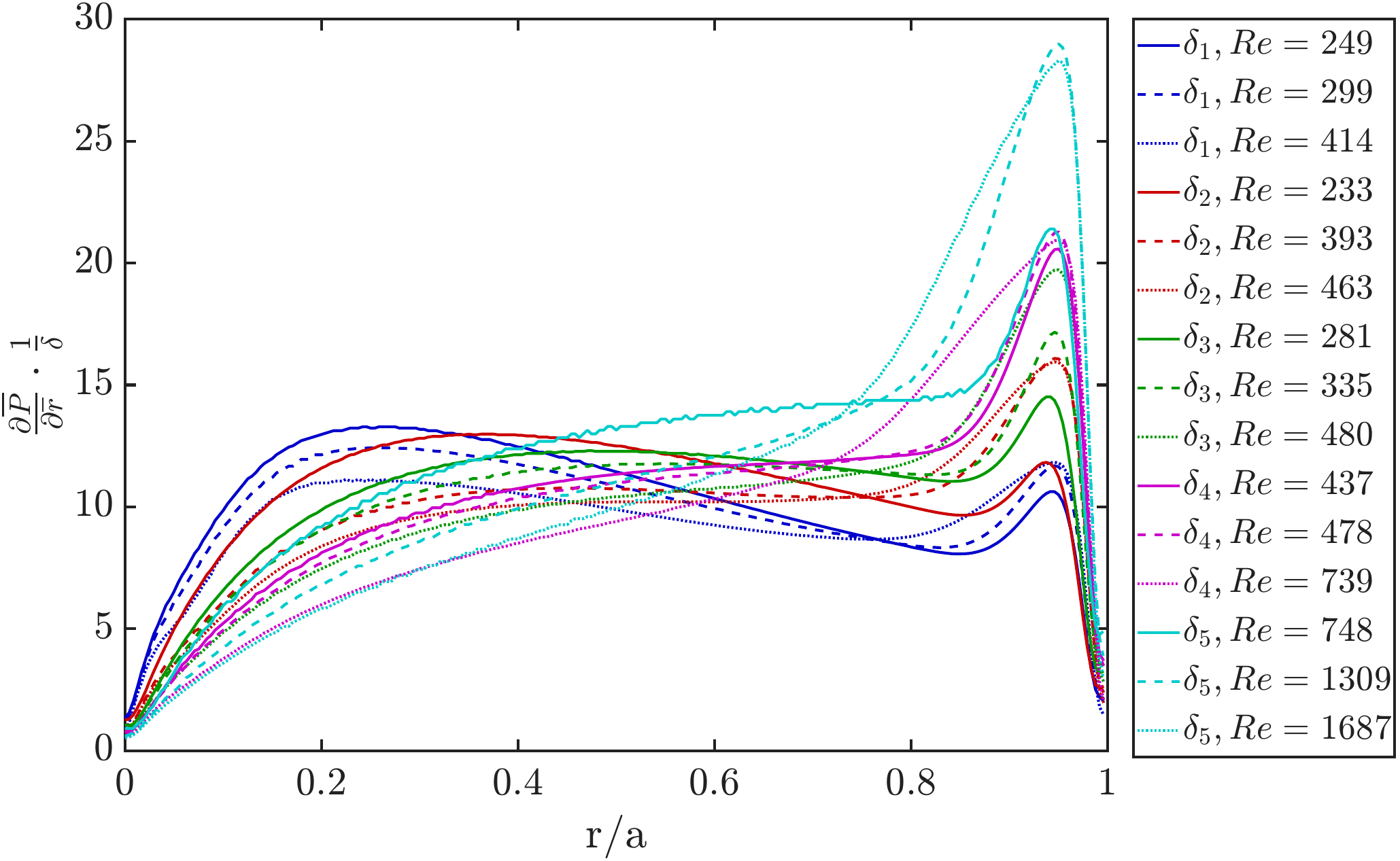}%
        \label{fig:dPdr_mod_De}}
    
    \caption{Dimensionless radial pressure gradient $\dfrac{1}{\delta}\dfrac{\partial \overline{P}}{\partial \overline{r}}$ along the mid-plane $\overline{z} = 0.5$ for (a) small Dean numbers and (b) moderate Dean numbers. }
    \label{fig:dPdr}
\end{figure}

Fig. \ref{fig:dPdr} displays the radial pressure gradient along the radial direction. This figure confirms that the pressure gradient is positive for all curvature ratios and Dean numbers. Note that the pressure is scaled by the dynamic pressure, whereas the differential radius is scaled by the channel width. The dimensionless radial pressure gradient is then multiplied by $1/\delta$ in order to compare profiles from different curvatures. 
At high curvature ratio and/or low Dean number, the radial pressure gradient, like the streamwise velocity profile (Figs. \ref{fig:mat3x3} and \ref{fig:Wscaled_d4_d5}), exhibits an extremum near the inner wall. This is not surprising since the centrifugal force acts as a source term in the momentum conservation equation, and the pressure gradient and viscous (resp. inertial) terms follow similar evolution in the cross-section. As the curvature ratio decreases and the Dean number increases, the local extrema shifts toward the outer wall, which is very similar to the shift of the peak of the streamwise velocity. {\color{black}Note that the numerical noise observed at the smallest curvature ratio is associated with the cell elongation in the azimuthal direction at $\delta_5$ $(\frac{R_5 d\theta}{dz} \approx 30 )$}. Using a finer mesh, allows to damp the numerical noise while the pressure gradient profile remains unchanged.

\subsection{Instabilities at high Dean number} \label{sec:unstableFlow}
\begin{figure}[ht]
    \centering
    \includegraphics[width=0.75\linewidth]{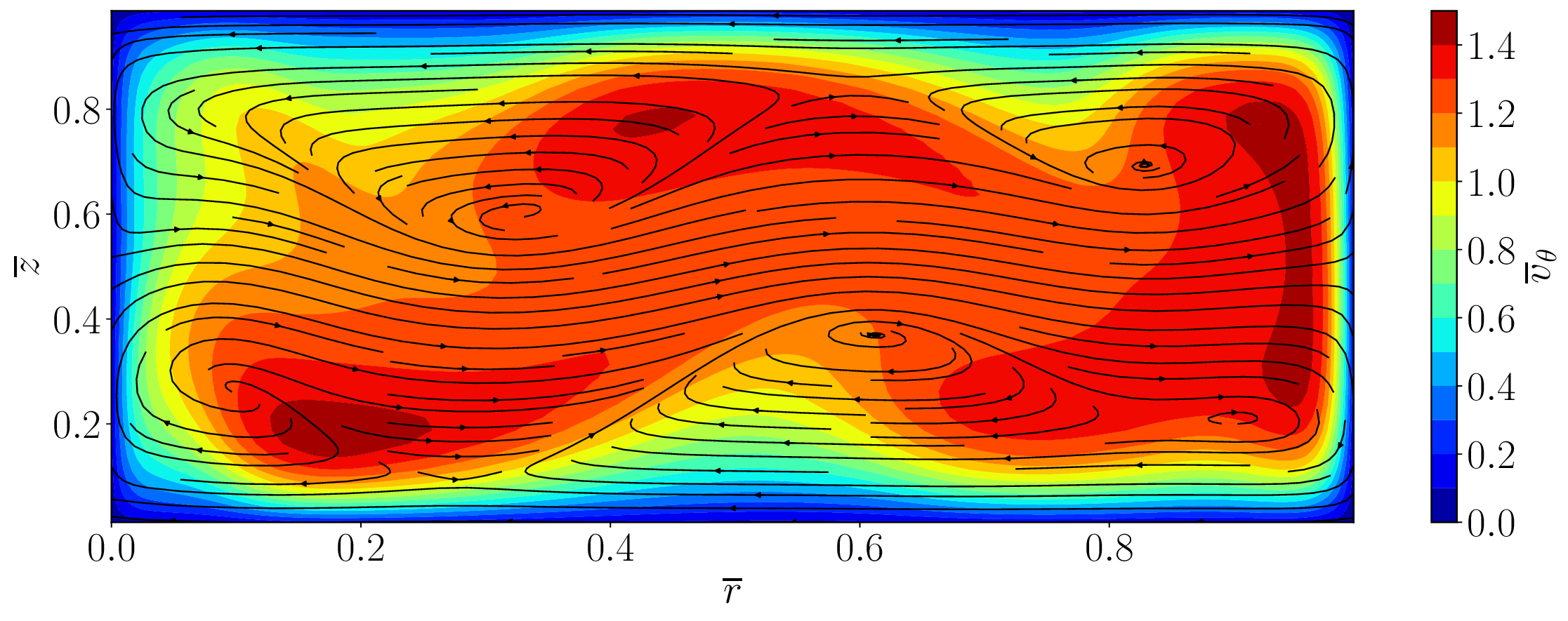}
    \caption{Contours of the azimuthal velocity $\overline{v}_\theta \text{ for } De = 220 $ }
    \label{fig:contour_inst}
\end{figure}

In all the results shown in this work, the flow remained steady while traveling for at least $10\pi$. At high Dean numbers and high curvature ratio, the flow is not unconditionally stable. Indeed, when flow inertia is relatively strong, perturbation of numerical nature grows in time without being dissipated by viscous effects. Fig.\ref{fig:contour_inst} shows an example ($\mathrm{De} \gtrsim 200$ and $\delta_1$) where the flow becomes unstable once it has traveled, on average, an angle of $\theta \sim 20\pi$. This instability takes the form of an elongated wave propagating in the cross-section similarly to what has been described by \cite{matsson1990} in the case of rotating curved channel flow. In their case, the Coriolis force comes into play, in addition to the centrifugal force, and depending on the direction of the rotation of the flow, the instabilities due to Coriolis and centrifugal effects interact to either cancel or enhance each other \cite{matsson1990}.
It is interesting to observe similar motion of the transient structures even in the absence of Coriolis forces.

\newpage

\bibliography{references}

\end{document}